\documentclass[twocolumn,usenatbib]{aastex631}

\usepackage{xifthen}
\usepackage{color}
\usepackage{multirow}
\usepackage{amsmath}

\usepackage[thinc]{esdiff}  

\usepackage{txfonts}

\usepackage{xfrac}

\newcommand{\thard}{\tau}
\newcommand{\hardrchar}{a_c}
\newcommand{\hardainit}{a_\tr{init}}
\newcommand{\hardaisco}{a_\tr{isco}}
\newcommand{\hardnuinner}{\nu_\tr{inner}}
\newcommand{\hardnuouter}{\nu_\tr{outer}}
\newcommand{\harddadtnorm}{H_a}

\newcommand{\holodeck}{\texttt{holodeck}}

\newcommand{\mbh}{M_\textrm{BH}}
\newcommand{\mbulge}{M_\mathrm{bulge}}
\newcommand{\mmbulge}{{$\mbh$--$\mbulge$}}
\newcommand{\mmbulgefunc}{\mbh}
\newcommand{\mmbulgefbulge}{f_{\star,\tr{bulge}}}
\newcommand{\gsmffunc}{\Psi}
\newcommand{\bhhost}{SMBH--host}
\newcommand{\tobs}{T_{\rm obs}}
\newcommand{\tgal}{T_\tr{gal-gal}}

\definecolor{purple1}{rgb}{0.6, 0.0, 0.8}
\definecolor{green1}{rgb}{0.25, 0.5, 0.25}
\definecolor{red1}{rgb}{0.7, 0.15, 0.15}

\newcommand*{\needcite}[1]{
    \ifthenelse{\equal{#1}{}}{
        {\color{red1}[?]}
    }{
        {\color{red1}[#1]}
    }
}

\newcommand{\tr}[1]{\textrm{#1}}

\newcommand{\msol}{\tr{M}_{\odot}}

\newcommand{\yr}{\mathrm{yr}}        
\newcommand{\pyr}{\textrm{yr}^{-1}}

\newcommand{\ndens}{\eta}
\newcommand{\ndensgalgal}{\eta_\tr{gal-gal}}
\newcommand{\nstardens}{\eta_\star}

\newcommand{\mstar}{m_{\star1}}
\newcommand{\mstarsec}{m_{\star2}}
\newcommand{\qstar}{q_\star}

\newcommand{\hc}{h_\tr{c}}
\newcommand{\hs}{h_\tr{s}}
\newcommand{\hscirc}{h_\tr{s,circ}}

\newcommand{\mchirp}{\mathcal{M}}     

\newcommand{\distcom}{d_c}   

\newcommand{\lgw}{L_\tr{GW}}

\newcommand{\E}[1]{\times\nobreak10^{#1}}

\newcommand{\lr}[2][]{
    \ifthenelse{\equal{#1}{}}{
        {\left(#2\right)}
    }{
        {\left(#2\right)}^{#1}
    }
}

\newcommand{\lrs}[2][]{
    \ifthenelse{\equal{#1}{}}{
        {\left[#2\right]}
    }{
        {\left[#2\right]}^{#1}
    }
}

\newcommand{\scale}[3][]{
    \ifthenelse{\equal{#1}{}}{
        \lr{ \frac{#2}{#3} }
    }{
        {\lr[#1]{ \frac{#2}{#3} }}
    }
}
\newcommand{\scales}[3][]{
    \ifthenelse{\equal{#1}{}}{
        \lrs{ \frac{#2}{#3} }
    }{
        {\lrs[#1]{ \frac{#2}{#3} }}
    }
}

\newcommand{\libphenom}{\textit{Phenom}}
\newcommand{\libphenomuniform}{\textit{Phenom+Uniform}}
\newcommand{\libphenomastro}{\textit{Phenom+Astro}}
\newcommand{\libgwonly}{\textit{GWOnly}}
\newcommand{\libgwonlyuniform}{\textit{GWOnly+Uniform}}

\newcommand{\libphenomext}{\textit{Phenom-Ext}}
\newcommand{\libphenomextastro}{\textit{Phenom-Ext+Astro}}
\newcommand{\libgwonlyext}{\textit{GWOnly-Ext}}
\newcommand{\libgwonlyextastro}{\textit{GWOnly-Ext+Astro}}

\newcommand{\hdall}{\textit{HD-w/MP+DP+CURN}}
\newcommand{\hdgp}{\textit{HD-DMGP}}

\newcommand{\tlifetime}{\tau_f}

\newcommand{\gsmfNormTot}{\Psi_0}
\newcommand{\gsmfnorm}{\psi_0}
\newcommand{\gsmfnormz}{\psi_z}
\newcommand{\gsmfMassTot}{M_\psi}
\newcommand{\gsmfmass}{m_{\psi,0}}
\newcommand{\gsmfmassz}{m_{\psi,z}}
\newcommand{\mmbamp}{\mu}
\newcommand{\mmbplaw}{\alpha_\mu}
\newcommand{\mmbscatter}{\epsilon_\mu}

\usepackage{layouts}
\usepackage{natbib}
\defcitealias{nanograv_15yr_dataset}{NG15}
\defcitealias{nanograv_15yr_detchar}{NG15detchar}
\defcitealias{nanograv_15yr_gwb}{NG15gwb}
\defcitealias{nanograv_15yr_new_physics}{NG15newphys}

\shorttitle{NANOGrav 15 yr Astrophysical Interpretation}
\shortauthors{The NANOGrav Collaboration}

\begin{document}

\title{The NANOGrav 15 yr Data Set:\\Constraints on Supermassive Black Hole Binaries from the Gravitational Wave Background}

\author[0000-0001-5134-3925]{Gabriella Agazie}
\affiliation{Center for Gravitation, Cosmology and Astrophysics, Department of Physics, University of Wisconsin-Milwaukee,\\ P.O. Box 413, Milwaukee, WI 53201, USA}
\author[0000-0002-8935-9882]{Akash Anumarlapudi}
\affiliation{Center for Gravitation, Cosmology and Astrophysics, Department of Physics, University of Wisconsin-Milwaukee,\\ P.O. Box 413, Milwaukee, WI 53201, USA}
\author[0000-0003-0638-3340]{Anne M. Archibald}
\affiliation{Newcastle University, NE1 7RU, UK}
\author[0000-0003-2745-753X]{Paul T. Baker}
\affiliation{Department of Physics and Astronomy, Widener University, One University Place, Chester, PA 19013, USA}
\author[0000-0003-0909-5563]{Bence B\'{e}csy}
\affiliation{Department of Physics, Oregon State University, Corvallis, OR 97331, USA}
\author[0000-0002-2183-1087]{Laura Blecha}
\affiliation{Physics Department, University of Florida, Gainesville, FL 32611, USA}
\author[0000-0002-7001-0728]{Alexander Bonilla}
\affiliation{Observatório Nacional, Rua General José Cristino 77, São Cristóvão, 20921-400 Rio de Janeiro, RJ, Brazil}
\affiliation{Facultad de Ciencias, Departamento de Matem\'aticas, Universidad El Bosque, Ak. 9 \# 131 A - 2, Bogot\'a, Colombia}
\author[0000-0001-6341-7178]{Adam Brazier}
\affiliation{Cornell Center for Astrophysics and Planetary Science and Department of Astronomy, Cornell University, Ithaca, NY 14853, USA}
\affiliation{Cornell Center for Advanced Computing, Cornell University, Ithaca, NY 14853, USA}
\author[0000-0003-3053-6538]{Paul R. Brook}
\affiliation{Institute for Gravitational Wave Astronomy and School of Physics and Astronomy, University of Birmingham, Edgbaston, Birmingham B15 2TT, UK}
\author[0000-0003-4052-7838]{Sarah Burke-Spolaor}
\affiliation{Department of Physics and Astronomy, West Virginia University, P.O. Box 6315, Morgantown, WV 26506, USA}
\affiliation{Center for Gravitational Waves and Cosmology, West Virginia University, Chestnut Ridge Research Building, Morgantown, WV 26505, USA}
% ...

\collaboration{1000}{The NANOGrav Collaboration}
\noaffiliation

\correspondingauthor{The NANOGrav Collaboration}
\email{{comments@nanograv.org}}

\begin{abstract}
The NANOGrav 15 yr data set shows evidence for the presence of a low-frequency gravitational-wave background (GWB).  While many physical processes can source such low-frequency gravitational waves, here we analyze the signal as coming from a population of supermassive black hole (SMBH) binaries distributed throughout the Universe.  We show that astrophysically motivated models of SMBH binary populations are able to reproduce both the amplitude and shape of the observed low-frequency gravitational-wave spectrum.  While multiple model variations are able to reproduce the GWB spectrum at our current measurement precision, our results highlight the importance of accurately modeling binary evolution for producing realistic GWB spectra.  Additionally, while reasonable parameters are able to reproduce the 15 yr observations, the implied GWB amplitude necessitates either a large number of parameters to be at the edges of expected values, or a small number of parameters to be notably different from standard expectations.  While we are not yet able to definitively establish the origin of the inferred GWB signal, the consistency of the signal with astrophysical expectations offers a tantalizing prospect for confirming that SMBH binaries are able to form, reach sub-parsec separations, and eventually coalesce.  As the significance grows over time, higher-order features of the GWB spectrum  will definitively determine the nature of the GWB and allow for novel constraints on SMBH populations.
\end{abstract}



\section{Introduction}
\label{sec:intro}

Strong observational evidence suggests that most, if not all, massive galaxies contain supermassive black holes (SMBHs) at their centers \citep{1998Natur.395A..14R}. Additionally, hierarchical structure formation causes frequent galaxy mergers \citep{1977ApJ...217L.125O, 1980MNRAS.191P...1W, 1993MNRAS.262..627L}, naturally leading to the formation of SMBH binaries, which may also merge \citep{BBR1980, Milosavljevic2001}.  At the last stages of their evolution, these binaries produce strong nanohertz gravitational wave (GW) emission that can be targeted by pulsar timing arrays (PTAs), which systematically monitor a large number of millisecond pulsars. By detecting coherent deviations in the times of arrival of pulsar signals, PTAs can observe a stochastic gravitational wave background (GWB) from the superposition of many unresolved binaries, as well as individually resolved sources on top of the background \citep{2019A&ARv..27....5B, 2021arXiv210513270T}.

The North American Nanohertz Observatory for Gravitational Waves (NANOGrav) 12.5-year data set showed evidence of a common-spectrum red noise process consistent with a GWB \citep{NANOGrav12p5_background}. This result was confirmed by the Parkes Pulsar Timing Array \citep[PPTA;][]{Goncharov21}, the European Pulsar Timing Array \citep[EPTA;][]{Chen21}, and the International Pulsar Timing Array \citep[IPTA;][]{IPTA_2022}.  The NANOGrav 15 yr data set shows that the common uncorrelated red noise (CURN) signal discovered in \citet{NANOGrav12p5_background} persists with greater significance and is now detected in a larger number of pulsars \citep[][{\it hereafter} \citetalias{nanograv_15yr_gwb}]{nanograv_15yr_gwb}. Additionally, for the first time, there is evidence of inter-pulsar correlations following the characteristic Hellings-Downs (HD) pattern \citep{Hellings1983} expected for an isotropic GWB.  Careful analyses of the detection significance give false-alarm probabilities of $\approx 10^{-4} - 10^{-3}$ ($\approx 3\,\sigma$).

In this paper, we investigate whether the NANOGrav 15 yr results can be explained as a stochastic GWB produced by a cosmic population of SMBH binaries.  While SMBH binaries have long been expected to produce such a background, a wide variety of alternative models exist, many of which invoke new physics that departs from the standard model and $\Lambda$ Cold Dark Matter ($\Lambda$CDM) cosmology. We refer the reader to \citet[][\textit{hereafter} \citetalias{nanograv_15yr_new_physics}]{nanograv_15yr_new_physics} for an analysis of the NANOGrav 15 yr results in the context of new-physics models, such as cosmic inflation, scalar-induced GWs, domain walls, cosmic strings, and first-order phase transitions.

\subsection{The galaxy--SMBH connection}

Our understanding of galaxy formation and evolution has rapidly progressed in the last few decades. This includes the definitive and now direct observation of SMBHs in galaxy centers  \citep{Ghez1998, GRAVITY18, EHT_M87,EHT_SgrA}.  The mass of the central SMBH strongly correlates with global properties of the host galaxy (e.g., the stellar velocity dispersion of the galactic bulge, the bulge mass and luminosity), with tight correlations spanning several orders of magnitude in SMBH mass \citep{1989IAUS..134..217D, 1993nag..conf..197K, 1998AJ....115.2285M, 2000ApJ...539L..13G, 2002ApJ...574..740T, 2004ApJ...604L..89H, 2009ApJ...698..198G, KH13, MM13, 2016ApJ...818...47S}.  These trends strongly imply coordinated evolution between SMBHs and their host galaxies, which may be driven by a variety of mechanisms such as galaxy mergers, secular dynamics, stellar feedback, and feedback from active galactic nuclei \citep[AGN; ][]{2005Natur.433..604D, 2008ApJS..175..356H, 2008MNRAS.391..481S}. SMBHs are believed to play particularly significant roles in shaping the structure of massive galaxies \citep{2006MNRAS.365...11C, 2012ARA&A..50..455F, 2014MNRAS.444.1518V, 2015MNRAS.446..521S, 2017MNRAS.465.3291W}, but many fundamental aspects, such as the formation channels of SMBH seeds in the early Universe or how AGN feedback shapes the host galaxies, are still poorly constrained via observations. The relevant physical processes are also very difficult to model theoretically, as they span size scales from galaxies ($\sim 10$ kpc) to SMBH event horizons ($\sim 10^{-5}$ pc). Similar challenges limit our ability to directly model the process of SMBH binary formation and evolution.

\subsection{SMBH binary evolution}

The formation of SMBH binaries begins with the merger of two galaxies, each hosting a central SMBH. At different stages of the evolution of the SMBH pair, different physical processes dominate energy and angular momentum extraction, which drives the binary to closer separations (\citealt{BBR1980}; see\citealt{DeRosa2019} for a recent review).  Initially, the SMBHs are a gravitationally unbound pair (a dual SMBH) falling towards the center of the merging host \citep{Barnes+Hernquist-1992} via dissipative ``hardening'' processes, such as dynamical friction \citep{Chandrasekhar-1943, Antonini+Merritt-2012}.  Once the mass enclosed within the orbit is comparable to the mass of the binary (typically at $\sim$ parsec-scale separations), the two black holes become a gravitationally bound pair \citep[a SMBH binary;][]{Merritt+Milosavljevic-2005}. At these separations, the timescale for GW-driven inspiral is generally still longer than the Hubble time, and their GW frequencies are orders of magnitude below those that PTAs can probe.

The astrophysical environment of the binary is therefore crucial for bringing these systems to the PTA bands and ultimately to their final coalescence.  Scattering of individual stars that pass close to the SMBHs can extract energy and angular momentum from the system, hardening the binary orbit \citep{Yu2002}. In some cases, the supply of stars on close orbits may be insufficient, and the binary would fail to merge within a Hubble time \citep{BBR1980}. However, this so-called ``final-parsec problem'' has a number of potential theoretical solutions \citep[e.g.,][]{2006ApJ...642L..21B, Holley06, Khan11, Holley15}. Similarly, in gas-rich systems, circumbinary gas disks can also catalyze the binary evolution \citep{Escala2005, Dotti+2007, haiman09}, but the efficiency of this process or whether the gas pushes the binary inward or outward is still unclear \citep{Munoz2019, Moody2019, Duffell2020, Siwek+2023}.

If a binary stalls for longer than the time between successive galaxy mergers, a second galaxy could bring a third SMBH into the system. Triple SMBH interactions can greatly reduce the timescale for a SMBH binary merger and may also cause the ejection of the lightest SMBH from the system \citep{1974ApJ...190..253S, Volonteri2003, Hoffman2007, Bonetti2016, Bonetti2018a}. Once a SMBH binary reaches a sufficiently small separation the GWs will dominate its evolution, carrying away energy and angular momentum and leading the SMBHs to coalescence \citep{Peters+Mathews-1963}.


\subsection{Electromagnetic signatures of SMBH binaries and multi-messenger prospects}

Many studies have used electromagnetic observations of AGN to find candidate SMBH pairs and binaries \citep[for reviews, see:][]{Komossa-2006, Popovic-2012, DeRosa2019, 2022LRR....25....3B}. Dual AGN, i.e., galaxies with two unbound, actively accreting SMBHs, have been identified at $\gtrsim$ kiloparsec separations \citep[e.g.,][and references therein]{Koss+2012, 2022ApJ...925..162C}.  However, spatially resolving the two SMBHs becomes increasingly challenging as their separation decreases.  Spectroscopic features, such as the kinematic offset of AGN narrow lines, can also be used to identify AGN in merging galaxies \citep[e.g.][]{Comerford+2009, Comerford+Greene-2014}.  To date, only one parsec-scale pair has been confirmed with very long baseline interferometry \citep[VLBI;][]{Rodriguez2006,2017ApJ...843...14B} despite large-scale searches \citep{2011MNRAS.410.2113B, 2021ApJ...914...37B}.

Electromagnetic searches for sub-parsec SMBH binaries typically focus on features that encode the binary's orbital motion on the temporal or spectral variability of AGN.  Searches for offset broad emission lines have been used to identify several hundred candidates \citep{Tsalmantza+2011, Erac+2012, Shen_BLsearchI+2013, Ju+2013}, but this method is subject to false positives and other limitations \citep{2007ApJS..169..167G, runnoe15, Runnoe+2017, 2018ApJ...861...59P, 2021MNRAS.500.4065K}.  Periodically variable light curves \citep{farris14, dorazio15b, bowen18, DOrazioDiStefano:2018} have yielded a similar number of candidates \citep{Graham+2015b, Charisi+2016, LiuGezari+2019}, though these samples are likely to also suffer significant contamination \citep{Vaughan2016, Charisi+2018, 2018ApJ...856...42S, kelley19, xin20}.  Despite these challenges, the advent of large time domain surveys with the Vera Rubin observatory \citep{vro_overview}, combined with multi-wavelength observations and increasing PTA sensitivity to the GWB, offers exciting opportunities for deriving multiple independent constraints on SMBH populations \citep[e.g.][]{2019BAAS...51c.490K, 2022LRR....25....3B}. The prospects for low-frequency multimessenger astrophysics are discussed further in \S~\ref{sec:discussion}.

\subsection{The Astrophysical Imprint on the Gravitational Wave Background}

All of the binary inspiral processes discussed above are imprinted on the GWB created by a population of SMBH binaries. Therefore, studying the GWB constitutes an important channel to obtain significant and novel insights on galaxy and binary mergers.  For example, interactions with the binary environment and orbital eccentricities impact the shape of the GWB spectrum \citep{Sesana_2013b}.  Stellar- and gas-driven binary hardening will cause a flattening or turnover of the low-frequency GWB spectrum, relative to the single power law predicted for GW-only evolution \citep{Kocsis+Sesana-2011}.  The primary effect of eccentricity is to boost GW emission to higher frequencies, owing to the emission of GWs at higher harmonics beyond twice the binary orbital frequency, which dominates for circular orbits \citep{Enoki+Nagashima-2007}. However, at extreme eccentricities ($\gtrsim 0.9-0.95$), close pericentric passages drive very rapid binary inspiral, leading to an overall attenuation of GWB amplitude at all frequencies \citep[e.g.,][]{Kelley+2017b}.

GW observations will also probe the history of SMBH mass growth.  The GWB depends strongly on the distribution of binary chirp masses, $\mathcal{M}$, given by
 \begin{equation}
     \label{eq:chirp_mass}
     \mchirp = \frac{\left(m_1 m_2\right)^{3/5}}{M^{1/5}} = M \frac{q^{3/5}}{\left(1 + q\right)^{6/5}},
 \end{equation}
where  $q \equiv m_2 / m_1 < 1$ is the binary mass ratio, $M = m_1 + m_2$ is the total binary mass, and  $m_1$ and $m_2$ are the masses of each SMBH.  As a result, the GWB is intimately related to the SMBH mass function through its dependence on the chirp mass, and in turn to the scaling relations of SMBH mass with host galaxy properties. These relations are well studied in the local Universe but are unconstrained at higher redshifts, and thus detailed studies of the GWB will provide a novel path to probing these relations.

Previous stochastic GWB constraints have been used to probe the SMBH binary population, by comparing with theoretical predictions for SMBH binary formation and evolution. All GWB results were strictly upper limits  until the NANOGrav 12.5-year data set, but the limits were still potentially constraining.  The constraints were especially informative when combined with electromagnetic observations of binary AGN candidates \citep{2018ApJ...856...42S, 2018MNRAS.481L..74H, 2018ApJ...863L..36I, 2020ApJ...900L..42N}.  After the PPTA upper limit at 2.8\,nHz \citep{2013Sci...342..334S}, it was first suggested that this ruled out a large range of SMBH-binary model space \citep{2015Sci...349.1522S}. However, \citet{2016MNRAS.455L..72M} showed that the upper limits were consistent with a wide variety of plausible astrophysical models and that, in general, upper limits alone would be relatively unconstraining until they were about an order of magnitude smaller. Subsequent work showed the importance of analyzing many pulsars and accounting for their red noise and systematic errors in solar-system ephemerides when establishing PTA upper limits \citep{NANOGrav11yrGWB, 2020ApJ...893..112V, 2020ApJ...890..108H, 2022ApJ...932..105J}. Since the 12.5-year NANOGrav data set showed evidence for a common red-noise process consistent with (but not unambiguously attributable to) GWs, the measurement was shown to be consistent with a population of SMBH binaries with reasonable properties \citep{2021MNRAS.502L..99M}.

\subsection{Astrophysical modeling of SMBH-binary GWB}

\begin{figure*}[tbh!]
    \centering
    \includegraphics[width=1\textwidth]{{{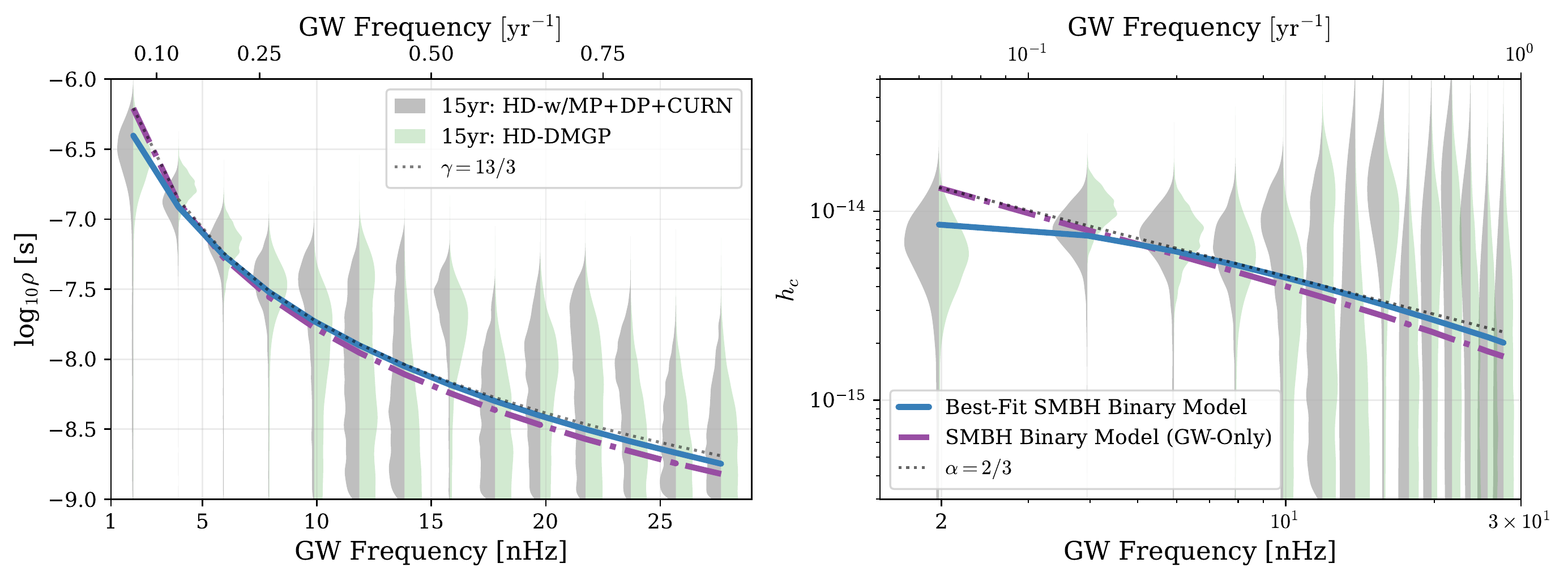}}}
    \caption{NANOGrav 15 yr GWB free-spectrum posteriors translated into the square root of timing residual power $(\rho$, left panel) and characteristic strain ($h_c$, right panel). The HD-correlated free-spectrum measured while simultaneously fitting for monopole-correlated (MP), dipole-correlated (DP), and uncorrelated red noise noise (CURN) free spectra ($\hdall$; gray violins, left-side) is compared against the $\hdgp$ model in which dispersion-measure variations are modeled using Gaussian Processes (green violins, right-side).  The black dotted lines show idealized power-law spectra ($\rho^2 \propto f^{-13/3}$ and $h_c \propto f^{-2/3}$) fit to the median posterior value for the amplitude obtained from the $\hdall$ model in \citetalias{nanograv_15yr_gwb}.  Over plotted are the best-fitting, simulated GWB spectra from models of SMBH binary populations produced in this analysis.  Two models are shown, one which includes environmentally driven binary evolution (blue) and another that assumes GW-only evolution (purple).  Both models are able to reproduce the data, while the environmentally driven model produces a slightly better fit.  We conclude that the observed GWB spectrum is consistent with astrophysically motivated expectations from populations of SMBH binaries.
    }
    \label{fig:15yr_spectrum}
\end{figure*}

Over the last few decades, a variety of different approaches have been used to model populations of SMBH binaries,\footnote{Throughout this work we will often use the term `SMBH binaries' to encompass SMBH pairs, even when the two SMBHs are not yet gravitationally bound but merely reside in the same galaxy.} with a wide range of predictions for the resulting GWB amplitude. (See Appendix \ref{sec:app_lit_models} for a summary of these model predictions and a comparison with the NANOGrav 15 yr results). Many of these studies start from either semi-analytic galaxy evolution models to obtain galaxy merger rates \citep{Rajagopal_1995} or halo merger-trees with added galaxies \citep{Menou+2001, Sesana+2004}, onto which a SMBH binary population model can be imposed.  In lieu of physically modeling environmentally driven SMBH binary evolution, galaxy mergers are often directly linked to the formation of a close SMBH binary emitting GWs at PTA frequencies, and a power-law form is assumed for the GWB \citep[e.g.,][]{Phinney-2001, Jaffe_2003, Wyithe_2003, Enoki_2004, Simon+Burke-Spolaor-2016}.  Some semi-analytic models also include prescriptions for physical processes that cause GWB spectra to deviate from a pure power law, such as interactions of the binary with the gaseous and stellar environment of its host galaxy, discreteness of the binary population, and orbital eccentricity \citep[e.g.,][]{Sesana+2008, Sesana_2009, Sesana_2013a, Ravi_2014, McWilliams_2014, Ryu_2018, Bonetti2018b, Chen_2020}.  Versions of the semi-analytic model approach have also been applied to catalogs of specific galaxies or quasars from observations \citep{Simon+2014, Rosado+Sesana-2014, Mingarelli+2017, Casey-Clyde+2022}.

An alternative to the semi-analytic modeling approach is to trace galaxy and SMBH evolution directly in cosmological hydrodynamics simulations \citep[e.g.,][]{Kulier_2015, Salcido+2016, Kelley+2017a, Kelley+2017b, Kelley+2018, Volonteri+2020, Siwek+2020, Curylo+Bulik-2022}. This approach has the advantage of providing detailed information about the internal structures of galaxies and how they interact with SMBHs via AGN fueling and feedback. However, cosmological hydrodynamical simulations are very computationally expensive compared to semi-analytic models, and even the highest-resolution simulations must rely on sub-grid prescriptions to model unresolved processes, including SMBH accretion, mergers, and feedback. Each of these complementary approaches therefore offers benefits and drawbacks, and importantly, each introduces certain systematics in their predictions for binary populations. In this work, we adopt a semi-analytic modeling approach to SMBH binary population synthesis and defer the use of cosmological hydrodynamics simulations for future work.

\subsection*{Summary \& Outline}
Figure~\ref{fig:15yr_spectrum} shows the GWB spectrum recovered from the 15 yr NANOGrav data, along with the best fitting simulated GWB spectra produced in this work.  In \S~\ref{sec:data} we summarize the NANOGrav 15 yr data set that forms the observational basis for this analysis, and the GWB spectra derived from it (grey and green `violins').  In \S~\ref{sec:methods}, we describe our methods of modeling populations of SMBH binaries and calculating the GWB spectra that they would produce.  There, we also detail the approach that we use to compare our simulations to the 15 yr data.  Our best-fitting models (colored curves) are presented in \S~\ref{sec:results}.

We find that astrophysically motivated models of SMBH binary populations are able to accurate reproduce the observed GWB spectrum (\S~\ref{sec:results_gwb_plaw}~\&~\ref{sec:results_binary_gwb}).  We focus our analysis on two population models.  One includes a self consistent prescription for environmentally driven binary evolution (blue), and the other assumes GW-only evolution (purple) which is still commonly used in the literature.  Both models are able to fit the data, while the environmentally driven case produces a slightly better match---particularly to the lowest frequency bin.  We present the binary evolution parameters favored by 15 yr spectra fits for both models (\S~\ref{sec:results_astro_posteriors}).  While the posterior distributions are broadly consistent with astrophysical expectations, parameters tend to be shifted towards values that produce larger GWB amplitudes than was previously most-favored.  Generally higher binary masses or densities, or highly efficient binary mergers are required to produce the observed amplitudes.  The characteristics of the implied binary populations are presented in \S~\ref{sec:results_astro_constraints}.

Our results are discussed in the context of the field in \S~\ref{sec:discussion}, along with highlights for the near future of low-frequency GW astronomy.

Throughout this paper we assume a WMAP9 cosmology with \mbox{$\Omega_m = 0.228$}, \mbox{$\Omega_b = 0.0472$}, and \mbox{$H_0 = 0.6933\ \mathrm{km\,s^{-1}\,Mpc^{-1}}$}.


\section{Pulsar Timing Array Data}
\label{sec:data}

This work is based on the NANOGrav 15 yr data set, which includes  68 pulsars, 67 of which have a baseline of at least 3 years and are included in the GWB analysis. The complete description of the data set can be found in \citet[][{\it hereafter} \citetalias{nanograv_15yr_dataset}]{nanograv_15yr_dataset}, while the detector characterization and noise modeling of individual pulsars is described in \citet[][{\it hereafter} \citetalias{nanograv_15yr_detchar}]{nanograv_15yr_detchar}. The detailed description of the Bayesian search for the GWB is presented in \citetalias{nanograv_15yr_gwb}. Here, we briefly summarize the measurement of the GWB spectrum from the NANOGrav data, focusing on the pieces which are necessary for the astrophysical interpretation presented in this paper.

PTA collaborations systematically monitor millisecond pulsars and record the times of arrival (TOAs) of their radio pulses. For each pulsar, a timing model is constructed, which estimates various factors affecting the TOAs including its astrometry (sky position, proper motion, and parallax), its spin period and spin period derivative, and binary parameters for pulsars with companions. Additionally, variations in the ionized interstellar medium along the line of sight, also known as the dispersion measure (DM), are included in our model. The analysis of each pulsar provides a best-fit estimate for the timing residuals, $r(t)$, which are the differences between the TOAs and the timing model. For more on the construction of the timing residuals in NANOGrav's data set, see \S~4 of \citetalias{nanograv_15yr_dataset}.

All red noise processes, including the GWB itself, are modeled with a Fourier basis computed on the TOAs, as discussed in \S~2 of \citetalias{nanograv_15yr_gwb}.  The frequencies are $f_i = i/\tobs$, where $\tobs = 16.03$~yr is the time between the first and last TOA included in this data set\footnote{This data set is named ``15 yr data set" since no single pulsar exceeds 16 years of observations, even though the total time spanned by the entire set of observations is $16.03$~yr.}. The search for a GW signal is performed by constructing the cross-correlations of residuals between pairs of pulsars, $a$ and $b$, i.e.,
\begin{equation}
    \langle r_{a}(t) r_{b}(t) \rangle \propto \int S_{ab}(f) df,
\end{equation}
where $f$ is the observer-frame GW frequency and $S_{ab}$ is the timing-residual cross-correlated power spectral density,
\begin{equation}
    S_{ab}(f) = \Gamma \left( \xi_{ab} \right) \Phi (f).
\end{equation}
Here, $\Phi (f)$ is the power spectral density (PSD) of the timing residuals describing the spectrum of the process that is common among all pulsars, and $\Gamma$ is the overlap reduction function, which describes the induced correlation between a pair of pulsars as a function of their angular separation, $\xi_{ab}$. The timing residual PSD is related to the characteristic GW strain, $h_c(f)$, by
\begin{equation}
\label{eq:dtpsd_to_hc}
\Phi (f) = \frac{h_c(f)^2}{ 12 \pi^2 f^3}.
\end{equation}
The overlap reduction function is given by $\Gamma \left( \xi_{ab} \right) = \delta_{ab}$ for a CURN model, and by the characteristic HD pattern in the case of an isotropic GWB \citep{Hellings1983}.

A specific spectral shape is typically prescribed to the common red noise process (see \S~\ref{sec:meth_binary_gws}).  Traditionally, a power law has been used, and the detailed spectral analysis presented in \citetalias{nanograv_15yr_gwb} shows support for this idealized, simple model. However, deviations appear at a variety of frequencies, which may skew the determination of a spectral slope (see Figure~\ref{fig:15yr_spectrum} and Figure~6 in \citetalias{nanograv_15yr_gwb}).  It is therefore important to model the individual Fourier coefficients independently rather than enforcing a specific spectral shape on the PSD. The resulting ``free spectrum" provides a minimally modeled Bayesian spectral characterization of PTA data. The free spectrum recovers the posterior of the common red-noise power spectrum at all sampling frequencies, and is parameterized by the coefficient $\rho$, where $\rho^2_i=\Phi(f_i)/\tobs$ is the power in the cross-correlated timing residuals.

Figure~\ref{fig:15yr_spectrum} shows the free-spectrum posteriors both in terms of the square-root, timing residual power ($\rho$) and converted into GW characteristic strain ($h_c$).  At the current signal-to-noise ratio, the spectral characterization of the signal is uncertain, and the recovered HD-correlated GWB signal may be impacted by non-HD-correlated noise. For our astrophysical interpretation, we adopt the free-spectrum posteriors from the 15 yr HD-correlated free-spectrum modeled simultaneously with additional monopole-correlated (MP), dipole-correlated (DP), and uncorrelated red noise (CURN).  This model, which we refer to as $\hdall$ (grey), provides the most conservative constraints on the recovered GWB spectrum.  As an additional comparison, we also analyze the HD-correlated free-spectrum posteriors utilizing an alternate model for DM variations, which we refer to as $\hdgp$ (green)---described in detail in \S~5.1 of \citetalias{nanograv_15yr_gwb}.

Figure~\ref{fig:15yr_spectrum} also shows the median posterior amplitude value for the idealized power-law fit to the $\hdall$ model.  Even though the power-law model provides an illustrative example for quick model comparison, we do not include it in the astrophysical interpretation, because it fails to encapsulate the full range of information contained in the free-spectrum posteriors.

The number of Fourier components used in an analysis is typically chosen based on the preference of the data for various red noise processes (e.g., in \citealt{NANOGrav12p5_background} the CURN model preferred only 5 frequencies, while in \citetalias{nanograv_15yr_gwb} that number has increased to 14).  While the CURN model prefers $14$ Fourier components in the 15 yr data set, the HD-correlated free spectra posteriors provide strong constraints only in the lowest $5$ frequency bins, and thus only those bins are used in this analysis. However, we find no difference in our results if we expand to using the full $14$ frequencies.


\section{Methods}
\label{sec:methods}

\begin{figure}
    \centering
    \includegraphics[width=0.8\columnwidth]{{{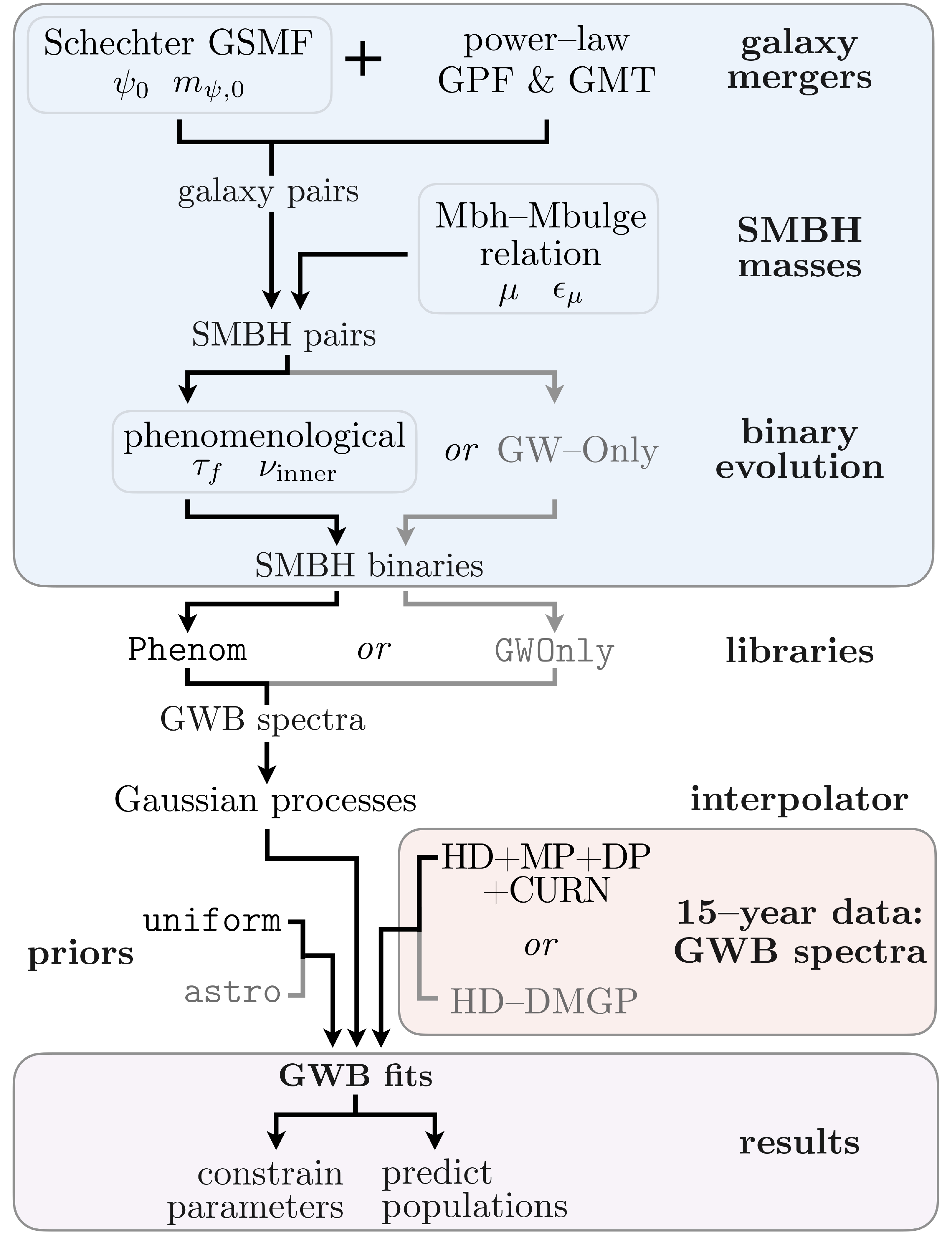}}}
    \caption{Schematic of the analysis pipeline used in this study.  Our population models are composed of galaxy mergers combined with SMBH masses and a prescription for binary evolution.  Galaxy merger rates are derived from the combination of a galaxy stellar-mass function (GSMF), galaxy pair fraction (GPF) and galaxy merger time (GMT).  Binary evolution can either follow the self-consistent, phenomenological approach, or it can assume GW-only evolution.  Libraries contain a large number of binary populations and their resulting GWB signals, which are calculated for varying uncertain physical parameters in both evolution scenarios.  Gaussian processes are used to interpolate across the library parameters when fitting against the NANOGrav 15 yr data.  Fits can be performed using broad, uniform priors or with more tightly constrained, astrophysically motivated priors.  Fits are performed against the 15 yr data using GWB spectra derived from the \hdall{} or \hdgp{} models (see text).}
    \label{fig:pipeline-schematic}
\end{figure}

Our goal is to constrain the properties of the underlying SMBH binary population that can produce a GWB consistent with the NANOGrav 15 yr data. Our approach consists of three main components described below, and depicted schematically in Figure~\ref{fig:pipeline-schematic}.

\paragraph{\bf{SMBH Binary Population Synthesis Simulations} (\S~\ref{sec:meth_binary_gws} -- \ref{sec:meth_libraries_params_effects})} We generate `libraries' of SMBH binary populations and their GW signals, exploring a large range of the binary formation/evolution parameter space.  For this, NANOGrav has developed a flexible framework for SMBH binary population synthesis called \holodeck{} \citep[][in prep]{holodeck},\footnote{\url{https://github.com/nanograv/holodeck}} which allows us to explore the binary population models and encompass systematic uncertainties.  Within \holodeck{}, we determine the number density of the cosmic population of SMBH binaries using semi-analytical models based on observationally constrained properties of galaxies and galaxy mergers.  Using a \bhhost{} relation, specifically the correlation between the mass of the SMBH and the mass of the stellar bulge, i.e. \mmbulge{}, we assign SMBH masses to the mergers and calculate the binary evolution from large separations down to the GW regime.  From each population, we compute the GWB signals they would produce.

\paragraph{\bf{Interpolation of the Population Synthesis Models} (\S~\ref{sec:interpolation})}  The simulated GWB spectra are sampled at discrete points of the multi-dimensional binary population parameter spaces that we explore.  We refer to the collection of simulated spectra for a given parameter space as a `library.'  We then use Gaussian processes (GPs) to interpolate between the population-synthesis simulations and predict the shape of the GWB spectrum for any point of the parameter domain.  This is necessary as the population simulations are too computationally expensive to run live while fitting against the NANOGrav data.

\paragraph{\bf{Fitting Population Synthesis Models Against PTA Data} (\S~\ref{sec:meth_ceffyl})} We use a Markov Chain Monte Carlo (MCMC) approach to fit the trained GPs against the input free-spectrum posteriors from \citetalias{nanograv_15yr_gwb}, generating posterior distributions of the binary population model parameters. From these, we constrain the different SMBH binary populations and evolutionary scenarios that could produce the observed \hbox{GWB}.

\subsection{GWs from SMBH binary populations}
\label{sec:meth_binary_gws}

The GWB spectrum can be calculated as the integrated GW emission of individual binaries throughout the Universe.  The characteristic strain of the GWB over a given logarithmic interval of frequency can be expressed as \citep{Phinney-2001, Wyithe_2003}
\begin{equation}
    \label{eq:gwb_mc}
    \hc^2(f) = \int dM dq dz \, \diffp{N}{{M} {q} {z} {\ln f_p}} \, \hs^2(f_p).
\end{equation}
The sky- and polarization-averaged GW spectral strain from a single, circular binary $\hs$ can be related to a binary's total GW luminosity, $\lgw$ as \citep{Finn+Thorne-2000}
\begin{align}
    \label{eq:strain_lum_circ}
    \hs^2(f) & = \frac{G}{c^3} \, \frac{\lgw}{\lr[2]{2\pi \,  f_p} \, \distcom^2} = \frac{32}{5 c^8} \,  \frac{\left(G\mchirp\right)^{10/3}}{\distcom^2} \lr[4/3]{2\pi f_p}.
\end{align}
Here, $\distcom$ is the comoving distance to a source at redshift $z$. Because the GW frequency is twice the orbital frequency for circular binaries, the observer-frame GW frequency $f$ can be related to the rest-frame orbital frequency $f_p$ as $f = 2 f_p / (1+z)$.  Throughout this paper, we take the chirp mass $\mchirp$ (Equation~\ref{eq:chirp_mass}), and by extension the total binary mass $M$, to be intrinsic rest-frame properties of the binary.

In practice, it is much more convenient to calculate a comoving volumetric number density of binaries $\ndens \equiv dN/dV_c$, and use this quantity to infer the full population \citep{Rajagopal_1995, Jaffe_2003, Sesana+2008}
\begin{equation}
\begin{split}
    \label{eq:number_density_to_number_frequency}
    \diffp{N}{{M} {q} {z} {\ln f_p}} & = \diffp{\ndens}{{M} {q} {z}} \diffp{t}{{\ln f_p}} \diffp{z}{{t}} \diffp{V_c}{{z}}, \\
    & = \diffp{{\ndens}}{{M} {q} {z}} \cdot \thard(f_p) \cdot 4 \pi c \, (1+z) \, \distcom^2.
\end{split}
\end{equation}
Here $\thard(f_p) \equiv f_p / (df_p/dt)$ is the binary hardening timescale, the rest-frame duration that a binary spends in a given logarithmic interval of frequency.  Equation~(\ref{eq:number_density_to_number_frequency}) connects redshift evolution to the time-evolution of binary sources over frequencies.  For a circular binary evolving purely due to GW emission, the rate of semi-major axis change and the hardening timescale are given by \citep{Peters-1964}
\begin{equation}
    \begin{split}
    \label{eq:gw_hard}
    \diffp{a}{t}\bigg|_\tr{gw} = & -\frac{64 \, G^3}{5 \, c^5} \frac{m_1 \, m_2 \, M}{a^3}, \\
    \thard_\tr{gw} \equiv & \diffp{t}{{\ln a}} = \frac{5}{96}\left(\frac{G\mchirp}{c^3}\right)^{-5/3} \left(2 \pi f_p \right)^{-8/3}.
    \end{split}
\end{equation}

Combining the above equations with the comoving volume of a light-cone \citep[e.g.,][]{Hogg-1999},
\begin{equation}
    \diffp{V_c}{z} \diffp{z}{t} = 4\pi \, c \, (1+z) \, \distcom^2,
\end{equation}
gives the idealized expression for a GWB produced by circular, GW-only driven SMBH binaries \citep{Phinney-2001}:
\begin{align}
    \label{eq:gwb_ideal}
    \hscirc^2(f) = & \frac{4 \pi}{3 c^2} \, \lr[-4/3]{2\pi f} \\ \nonumber
        & \int dM \, dq \, dz \, \diffp{\ndens}{{M}{q}{z}} \, \frac{\lr[5/3]{G \mchirp}}{\lr[1/3]{1+z}}.
\end{align}
This motivates the common expression for GWB spectra as a power law of the form,
\begin{equation}
    \label{eq:power_law_fit}
    \hc(f) = A_{\yr} \cdot \lr[-\alpha]{f / \yr^{-1}},
\end{equation}
where $A_{\yr}$ is the GWB amplitude referenced at a frequency of 1\,yr$^{-1}$, and in the idealized case, $\alpha=2/3$.  Because the timing-residual power spectral density of a GW signal is related to the characteristic GW strain by Equation \eqref{eq:dtpsd_to_hc}, this ideal, power-law form of the GWB can be expressed relative to a reference frequency $f_{\textrm{ref}}$ equivalently as,
\begin{equation}
   \Phi(f) = \frac{A^2}{12\pi^2} \left(\frac{f}{f_{\textrm{ref}}}\right)^{\!-\gamma} f_\textrm{ref}^{-3}.
   \label{eq:psd_plaw}
\end{equation}
Note that we have defined the power-law indices to be positive quantities such that $h_c \propto f^{-\alpha}$ and $\Phi \propto f^{-\gamma}$. The power-law indices are therefore related as $\gamma = 3 + 2\alpha$, such that the idealized, GW-only index is $\gamma = 13/3$.

Realistic GWB spectra can deviate substantially from a power law,  primarily due to the following three effects:\\

\paragraph{\bf{Interactions with the binary environment}} Astrophysical processes that extract energy and angular momentum from the binary (e.g., via stellar and gaseous interactions) can accelerate its frequency evolution relative to the GW-only evolution.  Therefore, any binary hardening via processes other than GW emission will necessarily result in an attenuation of the GWB compared to a purely GW-driven spectrum, as binaries spend less time emitting GWs in a given frequency interval.  This effect is particularly important at low frequencies ($f \ll 1 \, \pyr$) where binaries can more easily couple to their local galactic environments \citep{BBR1980, Kocsis+Sesana-2011}, and where GW emission is weaker.  In fact, coupling between SMBHs and their astrophysical environments is required for binaries to reach the PTA band within a Hubble time.  The question is thus whether the resulting flattening (or turnover) in the GWB spectrum occurs within the PTA  band or at frequencies too low to currently be accessible.

\paragraph{\bf{Discreteness of the binary population}} Equation~\eqref{eq:number_density_to_number_frequency} assumes a continuous distribution of SMBH binaries across the ($M$, $q$, $z$, $f$) parameter space.  At low frequencies ($f \lesssim 1 \, \pyr$), the hardening timescale is very long, and a large number of binaries contribute to the GWB---making this approximation valid.  At higher frequencies ($f \gtrsim 1 \, \pyr$), however, the hardening timescale becomes shorter and the typical number of binaries producing the bulk of the GWB energy in a given frequency bin approaches unity \citep{Sesana+2008}.  In this regime, a continuous distribution overestimates the GWB signal. Properly accounting for the finite number of sources in each frequency bin therefore results in a steeper GWB spectrum at high frequencies (Ibid.).  While a given overall amplitude of the GWB can be produced by either a larger number of lower-mass SMBH binaries or a smaller number of higher-mass binaries, these differences change the frequency at which discreteness becomes important. As a result, they change the location and severity of the high-frequency spectral steepening. \\

\paragraph{\bf{Orbital eccentricity}} Unlike circular binaries that emit GWs at exactly twice the orbital frequency, eccentric binaries emit GW energy at all integer harmonics.  This leads to GW energy being moved from lower frequencies to higher frequencies \citep{Enoki_2004}.  Additionally, smaller pericenter distances tend to increase the rate of binary inspiral.  These factors produce a variety of effects, including a spectral turnover at low frequencies, a flatter spectrum at higher frequencies, and a ``bump'' in between \citep{Enoki+Nagashima-2007, Sesana_2013b, Huerta_2015, 2017MNRAS.470.1738C, Kelley+2017b}.  However, for these effects to be substantial, very large eccentricities ($e\gtrsim 0.9$) are necessary at very small separations (well within the PTA band).\footnote{While recent results suggest that circumbinary accretion disks may drive moderate eccentricities ($e \sim 0.4-0.5$) in some systems \citep{Zrake+2021, dorazio21, Siwek+2023}, the effects are unlikely to be detectable in the GWB. Such processes could be more important for individually detectable GW signals---particularly rapidly accreting ones that may be promising multi-messenger sources.}  Since this is not expected to be the case, we restrict the current analysis to circular binaries.

These effects highlight the additional information encoded in the deviations of the GWB spectra from a pure power law and the importance of careful modeling of the binary population.  The above considerations also demonstrate the need for explicit integration of the binary evolution that includes environmental interactions, the discreteness of binaries, and their expected cosmic variance.


\subsection{SMBH Binary Population Synthesis}
\label{sec:meth_binary_pops}

Many previous works have constructed model populations of SMBH binaries and obtained predictions for the resulting GWB. These have produced a wide range of predictions for the GWB amplitudes, which are summarized in Table \ref{Table:gwb_predictions}. These SMBH binary population models generally involve either semi-analytic models or cosmological hydrodynamic simulations. In this work, we focus only on semi-analytic models and defer the exploration of binary populations from cosmological simulations to a future study.

In general, three key components are required for modeling the binary populations responsible for the GWB: (i)~galaxy masses and merger rates (\S~\ref{sec:meth_galaxy_mergers}), (ii)~SMBH masses based on a galaxy--host relationship (\S~\ref{sec:meth_mmbulge}), and (iii)~a binary evolution prescription (\S~\ref{sec:meth_binary_evo}).  We choose particular parameterizations for each of these components following \citet{Chen+2019}, described below, which are implemented in the \holodeck{} code.  A large number of free parameters are required for any such type of population synthesis calculation---more than can be meaningfully fit by the existing data. We therefore identify key parameters to vary in the models considered here and adopt standard literature values for the rest. In our analysis, we use these models to construct numerous different libraries of binary populations and we explore the impact of varying these key parameters on the resulting GWB spectra.

For the binary evolution, we consider libraries using both a phenomenological binary inspiral model (dubbed \libphenom{}) and a naive GW-only inspiral scenario (\libgwonly{}).  In both cases, the libraries vary two parameters that determine galaxy number density (the normalization $\gsmfnorm$ and turnover mass $\gsmfmass$) and two parameters describing the \mmbulge{} relationship  (the normalization $\mmbamp$ and intrinsic scatter $\mmbscatter$).  The \libphenom{} library includes two additional parameters describing the total binary lifetimes $\tlifetime$, and the binary hardening rate at small separations $\hardnuinner$.  These models and parameters are described in detail in the following sections.  Table~\ref{Table:sam_params} lists all of the parameters, giving fiducial values when they are fixed and the prior distributions for those which are varied.  Table~\ref{Table:models} summarizes our different libraries and which parameters are varied in each.  In Appendix~\ref{sec:app_constrained_priors}, we also compare against larger, ``extended" models in which additional parameters are varied (\libphenomext{} \& \libgwonlyext{}).

\subsubsection{The Galaxy Merger Rate}
\label{sec:meth_galaxy_mergers}

The number density of galaxy mergers ($\ndensgalgal$) can be expressed \citep{Chen+2019} in terms of a galaxy stellar-mass function (GSMF; $\gsmffunc$), a galaxy pair fraction (GPF; $P$), and a galaxy merger time (GMT; $\tgal$):
\begin{equation}
    \label{eq:sam_model_stellar}
    \diffp{\ndensgalgal}{{\mstar}{\qstar}{z}} = \frac{\gsmffunc(\mstar,z')}{\mstar \ln\! \lr{10}} \, \frac{P(\mstar,\qstar,z')}{\tgal(\mstar,\qstar,z')} \diffp{t}{{z'}}.
\end{equation}
This distribution is calculated in terms of the stellar mass of the primary galaxy $\mstar$, the stellar mass ratio (\mbox{$\qstar = \mstarsec/\mstar \leq 1$}), and the redshift $z$.  Because the galaxy merger spans a finite timescale ($\tgal$) and corresponding redshift interval, we distinguish between the initial redshift at which a galaxy pair forms ($z' = z'[t]$ at some initial time $t$) and the redshift at which the system becomes a post-merger galaxy remnant ($z = z[t + \tgal]$).  An additional delay is required for binaries to reach the PTA frequency band, which is characterized in \S~\ref{sec:meth_binary_evo}.

The GSMF is defined as
\begin{equation}
\gsmffunc(\mstar, z') \equiv \diffp{\nstardens(\mstar, z')}{{\log_{10}{\mstar}}},
\end{equation}
i.e., the differential number-density of galaxies per decade of stellar mass.  The implementation used in this analysis described the GSMF in terms of a single Schechter function \citep{Schechter-1976}:
\begin{equation}
\label{eq:gsmf_schechter}
\gsmffunc(\mstar, z) = \, \ln\lr{10} \, \gsmfNormTot \cdot \scales[\alpha_\psi]{\mstar}{\gsmfMassTot} \exp\lr{-\frac{\mstar}{\gsmfMassTot}},
\end{equation}
where we have introduced $\gsmfNormTot$, $\gsmfMassTot$, and $\alpha_\psi$ as new variables. In order to allow the GSMF to vary with redshift, we parameterize these quantities as
\begin{equation}
\begin{split}
\log_{10}\lr{\gsmfNormTot / \tr{Mpc}^{-3}} = & \, \psi_0 + \psi_z \cdot z, \\
\log_{10}\lr{\gsmfMassTot / \msol} = & \, m_{\psi,0} + m_{\psi,z} \cdot z, \\
\alpha_\psi = & \, 1 + \alpha_{\psi,0} + \alpha_{\psi,z} \cdot z,
\end{split}
\end{equation}
such that each of these quantities has a simple linear scaling with redshift. This introduces six new dimensionless parameters into our models, corresponding to the normalization ($\gsmfnorm$, $\gsmfmass$, \& $\alpha_{\psi,0}$) and slope ($\gsmfnormz$, $\gsmfmassz$, \& $\alpha_{\psi,z}$) of the redshift scaling. In all of the analysis presented here, the latter three are always kept fixed at the fiducial values specified in Table~\ref{Table:sam_params}. The GSMF normalization and characteristic mass parameters $\gsmfnorm$ and $\gsmfmass$ are allowed to vary in our fiducial \libphenom{} library, while $\alpha_{\psi,0}$ is additionally varied in \libphenomext{}.

The GPF and GMT are defined as
\begin{eqnarray}
P(\mstar, \qstar, z') \equiv & \diffp{}{\qstar} \scale{N_{\star,\tr{pairs}}(\mstar, \qstar, z')}{N_\star(\mstar, z')}, \\
\tgal(\mstar,\qstar,z') \equiv & {\displaystyle \int_{a_{\star,i}}^{a_{\star,f}}} \lrs{\dot{a}_\star(\mstar,\qstar,z')}^{-1} da_\star,
\end{eqnarray}
where $\dot{a}_\star$ denotes the rate at which the merging galaxies' separation decreases. The GPF describes the number of observable galaxy pairs relative to the number of all galaxies.  The GMT is the duration over which two galaxies can be discernible as pairs, from an initial separation $a_{\star,i}$ at which they are associated with one another, until a final separation $a_{\star,f}$ after which they are no longer distinguishable as separate galaxies. These two distributions are typically determined empirically based on the detection of galaxy pairs in observational surveys and thus depend on observational definitions and selection criteria \citep[e.g.,][]{Conselice+2008, Mundy+2017, Snyder+2017, Duncan+2019}.

In practice, we parameterize $P(\mstar, \qstar, z')$ and $\tgal(\mstar, \qstar, z')$ as redshift-dependent power laws of $\mstar$, $\qstar$ and $z$ following \citet{Chen+2019}:
\begin{equation}
\begin{split}
    P(\mstar, \qstar, z') = & \, P_0 \, \scale[\alpha_p]{\mstar}{10^{11} \, \msol} \, \lr[\beta_p]{1 + z} \, q^{\gamma_p} \\
    \alpha_p = & \, \alpha_{p,0} + \alpha_{p,z} \cdot z \\
    \gamma_p = & \, \gamma_{p,0} + \gamma_{p,z} \cdot z.
\end{split}
\label{eq:gpf}
\end{equation}
\begin{equation}
\begin{split}
    \tgal(\mstar, \qstar, z') = & \, T_0 \, \scale[\alpha_t]{\mstar}{10^{11} \, \msol/h} \, \lr[\beta_t]{1 + z} \, q^{\gamma_t} \\
    \alpha_t = & \, \alpha_{t,0} + \alpha_{t,z} \cdot z \\
    \gamma_t = & \, \gamma_{t,0} + \gamma_{t,z} \cdot z.
\end{split}
\label{eq:gmt}
\end{equation}
As shown in Table \ref{Table:sam_params}, the values of the corresponding parameters are kept fixed to standard literature values in our fiducial model (\libphenom{}), but parameters governing the scaling of the GPF and GMT with redshift ($\beta_{p,0}$ \& $\beta_{t,0}$), the scaling of the GPF with mass ratio ($\gamma_{p,0}$), and the GMT normalization ($T_0$) are allowed to vary in our extended models, as described in more detail below.

\subsubsection{The SMBH--Host Relation}
\label{sec:meth_mmbulge}

The number density of galaxy mergers given in Equation~(\ref{eq:sam_model_stellar}) is a distribution that describes the number of galaxy pairs as a function of galaxy properties.  We assume a one-to-one correspondence between galaxy pairs and SMBH binaries and adopt a \bhhost{} relationship to translate from galaxies to SMBHs.  In this analysis, we restrict ourselves to the \mmbulge{} relationship, which relates the galaxy stellar bulge mass to the SMBH mass for each component of the binary as \citep{Marconi+Hunt-2003}
\begin{equation}
\log_{10}\! \lr{\mbh/\msol} = \mmbamp + \mmbplaw \log_{10}\!\scale{\mbulge}{10^{11} \, \msol} + \mathcal{N}\lr{0, \mmbscatter}.
\end{equation}
Here $\mathcal{N}(0, \mmbscatter)$ denotes normally distributed random scatter with a mean of zero and standard deviation of $\mmbscatter$ (in dex).  This relation depends on three model parameters that are allowed to vary in our analyses: the dimensionless BH mass normalization ($\mu$), the intrinsic scatter ($\epsilon_\mu$, in dex), and the power-law index $\alpha_\mu$ (which is varied only in our extended models, and is dimensionless). A fraction of the galaxy stellar mass is in the stellar bulge component ($\mbulge =  \, \mmbulgefbulge \cdot m_\star$), which we take to be $f_{\star,{\rm bulge}} = 0.615$ based on empirical bulge fraction measurements of massive galaxies from \citet{Lang+2014} and \citet{Bluck+2014}.  Using the \mmbulge{} relationship, we transform the number density of galaxy mergers to a number density of SMBH binaries via
\begin{equation}
    \label{eq:sam_model_bh}
    \diffp{\ndens}{{M}{q}{z}} = \diffp{\ndensgalgal}{{\mstar}{\qstar}{z}} \: \diffp{\mstar}{M} \: \diffp{\qstar}{q}.
\end{equation}

Equation~(\ref{eq:sam_model_bh}) provides an expectation value for the number of binaries in a point ($M$, $q$, $z$) in parameter space.  To discretize the SMBH binary population, and also measure the effects of cosmic variance, we assume that the true number of binaries in any given spatial volume is Poisson-distributed.  We then integrate the differential number of binaries over finite bins of parameter space to obtain the expected number of binaries in each bin. We generate multiple realizations by drawing many times from a Poisson distribution ($\mathcal{P}$) centered at that value. Finally, we sum over parameter-space bins to calculate the resulting GWB spectrum.  In practice, we implement Equation~(\ref{eq:gwb_mc}) as:
\begin{equation}
    \hc^2(f)\!=\!\!\!\sum_{M,q,z,f}\!\!\mathcal{P} \, \lr{\diffp{N}{{M} {q} {z} {\!\ln f_p}} \Delta M \, \Delta q \, \Delta z \, \Delta
\!\ln f} \, \frac{\hs^2(f_p)}{\Delta\!\ln f}.
\end{equation}

\subsubsection{Binary Evolution}
\label{sec:meth_binary_evo}

The final component for constructing the GWB is the most uncertain: the binary evolution from the initial galaxy merger until the eventual SMBH coalescence. Typically, interactions with the astrophysical environment (i.e., stars and gas in the host galaxy) are required to bring an SMBH binary into the PTA band within a Hubble time. For example, the high-mass binaries ($\mchirp \gtrsim 10^9\,\msol$) that dominate the GW signals in the PTA band must reach separations of $\sim 0.1\,\tr{pc}$ before GW emission becomes dominant and drives efficient inspiral.  Those binaries enter the NANOGrav band (currently: $\sim 1/15\,\rm yr^{-1} \approx 2\,\tr{nHz}$) when they reach separations of $\approx 0.05 \, \tr{pc}$ -- only a factor of two smaller.  This immediately implies that the environmental processes may play a non-negligible role in binary evolution, even after the binaries reach the NANOGrav band.

Even if environmental hardening is effective in bringing binaries to the PTA-detectable frequencies, binary lifetimes can still be many billions of years, and a large fraction of binaries may stall \citep{Kelley+2017a}.  This can lead to binaries reaching the PTA band at substantially lower redshifts than those at which their respective galaxy mergers occurred.

Detailed modeling of environmentally driven binary evolution can introduce dozens of free parameters, even when SMBH and galaxy parameters are known \textit{a priori}.  Ultimately, many of these parameters become significantly degenerate in determining the resulting shape of the GWB spectrum and the properties of the SMBH binaries producing it.  For this reason, we focus this analysis on a `phenomenological' model that is designed to capture the overall effects of more explicit binary evolution, while introducing only a small number of free parameters.  In these models, the hardening timescale is parameterized in terms of the evolution of the binary semi-major axis $a$ as
\begin{equation}
    \label{eq:hard_phenom}
    \frac{d a}{d t}\bigg|_\tr{phenom} = \harddadtnorm \cdot \lr[1-\hardnuinner]{\frac{a}{\hardrchar}} \cdot \lr[\hardnuinner-\hardnuouter]{1 + \frac{a}{\hardrchar}}.
\end{equation}
The hardening timescale is thus a double power law, with a break at the critical separation $\hardrchar$, and asymptotic behaviors of:
\begin{equation}
\frac{dt}{d\ln a}(a \ll \hardrchar) \sim a^{\hardnuinner},
\end{equation}
in the `inner' (small-separation) regime, and
\begin{equation}
\frac{dt}{d\ln a}(a \gg \hardrchar) \sim a^{\hardnuouter},
\end{equation}
in the `outer' (large-separation) regime.  Hardening rates are added linearly, such that the total rate of evolution when also including GW emission (Equation~\ref{eq:gw_hard}) is given by $da/dt = \lrs{da/dt}_\tr{phenom} + \lrs{da/dt}_\tr{GW}$. We assume a fixed value of $\hardnuouter=+2.5$ in all of our analysis, motivated by detailed literature models of dynamical-friction-driven evolution of SMBH binaries \citep{Kelley+2017a}. $\hardnuinner$, which controls the hardening rate of binaries as they approach and enter the PTA band, is allowed to vary in our models.

In addition to the two power-law indices ($\hardnuinner, \hardnuouter$), and the characteristic break separation ($\hardrchar$), the normalization ($\harddadtnorm$) is calculated such that the total lifetime of the binary matches a target $\tlifetime$, i.e.,
\begin{equation}
    \label{eq:hard_phenom_lifetime}
    \tlifetime = \int_{\hardainit}^{\hardaisco} \scale[-1]{da}{dt} da,
\end{equation}
where $\hardainit$ is the initial binary separation and \mbox{$\hardaisco \equiv 6 \, G M / c^2$} is the innermost stable circular orbit, where we consider the two SMBHs to have merged.  While this expression for $\hardaisco$ is based on the test-particle approximation (\mbox{$q \ll 1$}), the true value should differ by less than a factor of two \citep{Flanagan+Hughes-1998} for low SMBH spins, and the contribution to the total lifetime is always negligible for \mbox{$a \sim \hardaisco$}.  The total lifetime $\tlifetime$ is a key parameter that we vary in our models.

At numerous po14ints in our analysis we compare the self-consistent, phenomenological model (in the \libphenom{} and \libphenomext{} libraries) against a model where binaries decay only due to GW emission (\libgwonly{} and \libgwonlyext{} libraries).  In the GW-only model, we take the redshift (and thus source distance) to be the post-galaxy-merger redshift, without an additional delay, and set the binary evolution time in Equation~(\ref{eq:number_density_to_number_frequency}) to be that of GW-only evolution (i.e., Equation~\ref{eq:gw_hard}).  This model is not self-consistent, as GW-only evolution is unable to bring binaries to the PTA band within a Hubble time.  It is nonetheless a useful comparison, because the GW-only assumption is often still used in the literature and tends to produce the highest GWB amplitudes.

\begin{figure}
    \centering
    \includegraphics[width=1\columnwidth]{{{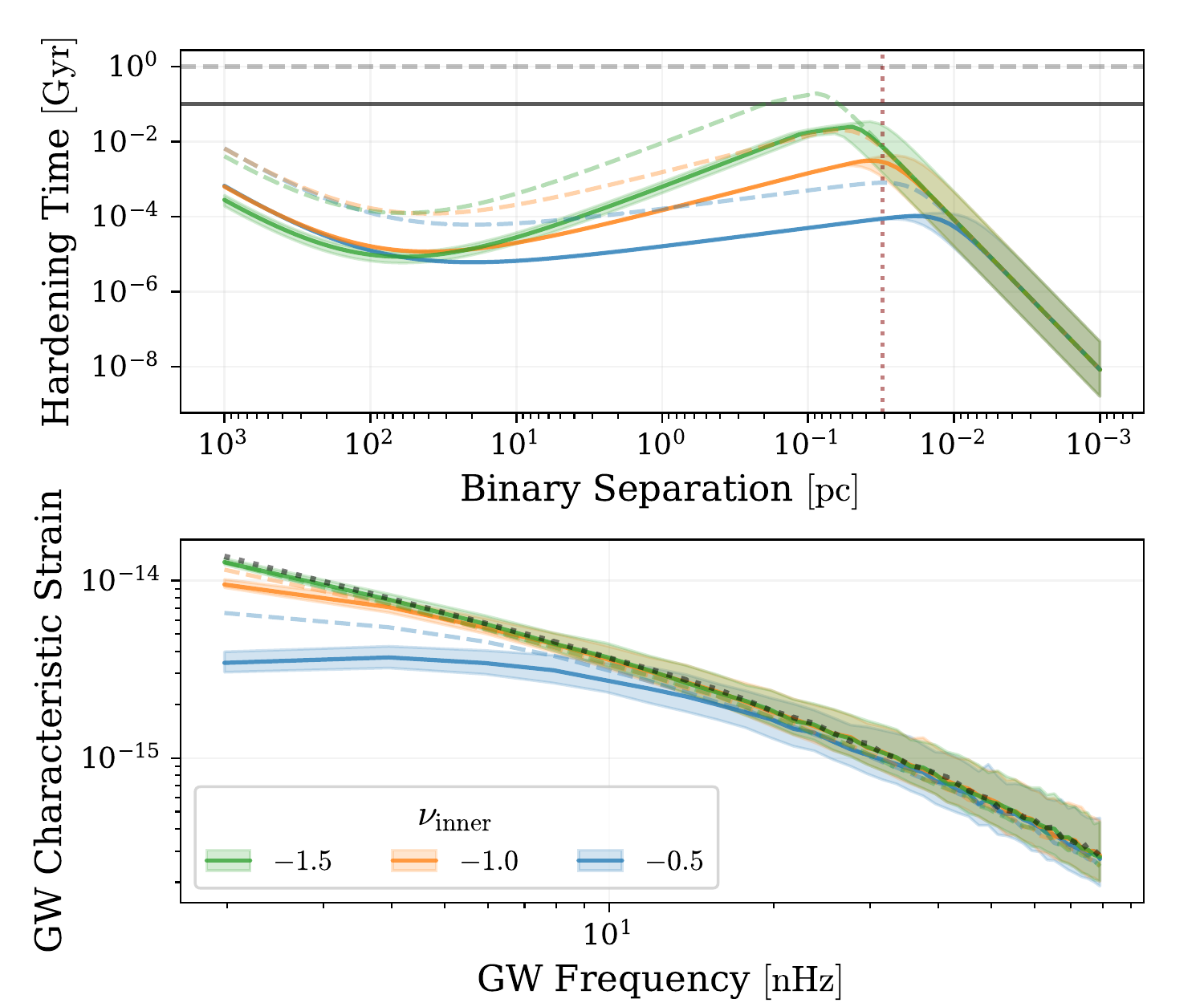}}}
    \caption{SMBH binary hardening timescales and GW spectra are shown for varying values of phenomenological binary evolution model parameters.  The \textit{top} panel shows the hardening time-scale ($\tau = dt/d\ln{}a$), with the black, horizontal lines corresponding to total binary lifetimes.  Solid lines correspond to binary lifetimes $\tlifetime = 0.1$~Gyr, while dashed are $1.0$~Gyr.  The small separation hardening rate power-law index, $\hardnuinner$, is also varied which changes the amount of time binaries spend at sub-parsec separations.  The vertical dotted line shows the separation at which an $M=10^9 \, \msol$ system reaches $f = 1/16.03\tr{ yr}$.  The \textit{bottom} panel shows the resulting GWB characteristic strain.  The dotted line shows GW-only driven evolution.  For clarity, the top panel includes only binaries with \mbox{$3\E{8} \, \msol < M \leq 3\E{9} \, \msol$} and \mbox{$0.1 < q \leq 1.0$}.  In both panels, the shaded regions denote the inter-quartile range.  Other parameters of these population are fiducial values for the \libphenom{} set of models (\S~\ref{sec:meth_libraries_params_effects}). Variation in the binary evolution parameters significantly impacts the shape and low-frequency amplitude of the GWB spectrum.}
    \label{fig:binary-evo}
\end{figure}

Figure~\ref{fig:binary-evo} shows the binary evolution and GWB spectra resulting from the phenomenological evolution model.  Total binary lifetimes of 0.1~and~1~Gyr are plotted with solid and dashed lines respectively, while varying small-separation power-law indices ($\hardnuinner$) are shown with different colors.  In each panel, the median and 50\% interquartile range of binaries are shown.  Note that in the \textit{top} panel, only binaries with \mbox{$3\E{8} \, \msol < M \leq 3\E{9} \, \msol$} and \mbox{$0.1 < q \leq 1.0$} are shown.  In the environmentally driven regime (larger separations), their hardening rate is determined such that their total lifetime matches the target value.  The narrow interquartile regions in the environmental regime reflect the small variations in hardening rate required to produce the target total lifetime, for this range of masses.

The GW hardening rate, which dominates at small separations, is determined entirely by the binary masses for our assumption of circular orbits.  In the phenomenological model, the hardening rate at larger separations is determined such that the total inspiral time matches the input binary lifetime. This means that shorter lifetime populations are forced to transition into the GW-driven regime at smaller separations.  The power-law indices also affect the transition point by determining which separations dominate the binary's evolution time.  More positive values of $\hardnuinner$ lead to flatter evolution trajectories with less and less time spent at sub-parsec binary separations.  For reference, the dotted vertical line shows the separation at which a binary with $M=10^9 \, \msol$ enters the 15 yr NANOGrav frequency band.  The two models with $\hardnuinner=-0.5$ and $\hardnuinner=-1.0$ lead to environmentally driven evolution until sufficiently small separations so that the resulting GWB spectral turnover (\textit{bottom} panel) is clearly visible in simulated 15 yr spectra.

We note that a value of $\hardnuinner=-1.0$ is well-motivated by numerical stellar scattering experiments of closely bound SMBH binaries \citep{sesana10, sesana15}. However, the true rate of environmental hardening for close binaries will depend on the stellar distribution in a given host galaxy, as well as on the role of gas-driven binary evolution, motivating the choice to allow $\hardnuinner$ to vary in our models.

The $\hardnuinner=-0.5$ variation produces a substantial attenuation of the GWB: up to a 50\% decrease in characteristic strain (75\% reduction in GW power).  Even though the $1.0$~Gyr lifetime model (dashed lines) qualifies as efficient and rapid binary evolution, the overall amplitude of the GWB at all frequencies is $\sim10\%$ lower than the GW-only model.  This is because a fraction of the binaries, specifically those which formed within a look-back time of $1.0$~Gyr, are unable to reach the PTA band before redshift zero, and thus do not contribute to the observable GWB.  This figure highlights that only in a narrow region of parameter space do realistic GWB spectra match predictions from GW-only models, but the differences between the self-consistent and GW-only models can be subtle.

\subsection{Libraries of SMBH Binary Populations and GWB Spectra}
\label{sec:meth_libraries_params_effects}

With the models described above, we use the \holodeck{} code to calculate libraries of SMBH binary populations and their resulting GWB spectra.  In each parameter space that we explore, we include a large number of sample points in the space in addition to many realizations of populations and spectra at each point.  We use the same GW frequency bins as the 15 yr NANOGrav data ($f_i = i / 16.03 \, \yr^{-1} = i\times 1.98~\tr{nHz}$, see \S~\ref{sec:data}) to calculate spectra.  Here we present the two primary libraries used in our analysis: \libphenom{} and \libgwonly{}, and outline some of their features.  In Table~\ref{Table:models} we summarize the parameters that are varied in each library, while the full list of model parameters (including fiducial values for fixed parameters and assumed prior distributions for varied parameters) is given in Table \ref{Table:sam_params}.


\begin{figure*}
    \centering
    \includegraphics[width=0.98\textwidth]{{{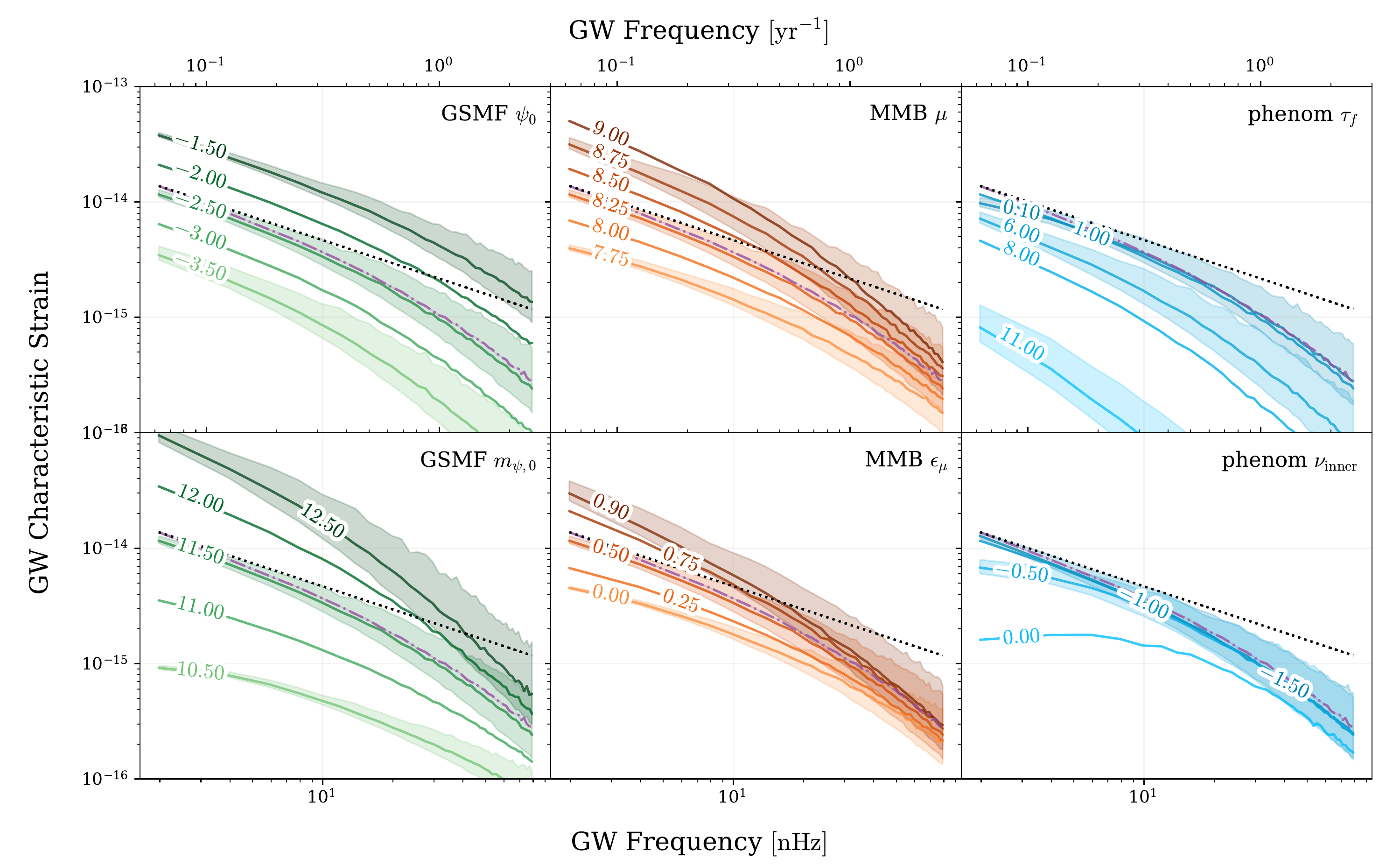}}}
    \caption{Comparison of GWB spectra for systematic variations in the parameters of our fiducial range of models (\libphenom{}).  In each panel, the indicated parameter is varied while the other parameters are fixed at typical values.  The purple dot-dashed line shows the spectrum from a fiducial GW-only model, while the black dotted line shows a pure $f^{-2/3}$ power law as a reference.  Variations in the GSMF parameters ($\gsmfnorm$---the dimensionless GSMF normalization, and $\gsmfmass$---the dimensionless GSMF turnover mass) are shown in green. Variations in the \mmbulge{} parameters ($\mmbamp$---the dimensionless \mmbulge{} normalization), and $\mmbscatter$---the \mmbulge{} scatter in dex) are in orange.  Variations in the SMBH binary lifetime ($\tlifetime$ in units of Gyr) and hardening power-law index ($\hardnuinner$) are shown in blue.  The numeric label indicates the value of the parameter for that particular curve, and the fixed values for all other parameters are: $\gsmfnorm = -2.5$, $\gsmfmass = 11.5$, $\mmbamp = 8.25$, $\mmbscatter = 0.5$ dex, $\tlifetime = 1.0$ Gyr, and $\hardnuinner = -1.0$.  The solid lines show the median for each parameter value from 10,000 realizations.  The shaded regions indicate the $16^{\mathrm{th}}$ and $84^{\mathrm{th}}$ percentiles of the distribution.  For clarity, we only plot the shaded regions for every other parameter value. Apart from the mostly degenerate mass parameters $\mmbamp$ and $\gsmfmass$, which nonetheless have a significant influence on the GWB spectrum, each of these parameters impacts the GWB spectra in distinct ways. This indicates the promise of GW observations for constraining the SMBH binary population.}
    \label{fig:gwb_anatomy}
\end{figure*}

Tens of free parameters are required in these models, many of which are poorly constrained either observationally or theoretically.  In addition, many of them are formally degenerate in their effects on the resulting GWB spectra \citep[e.g.,][]{Chen+2019}. For this reason, we adopt as our fiducial library, \libphenom{}, a model with six parameters $\{\gsmfnorm, \gsmfmass, \mmbamp, \mmbscatter, \tlifetime, \hardnuinner\}$ that produce SMBH binary populations and GWB spectra that effectively span the broader model uncertainties.  More specifically, this library varies the normalization and turnover mass of the GSMF ($\gsmfnorm$ and $\gsmfmass$,), along with the normalization and scatter of the \mmbulge{} relationship ($\mmbamp$ and $\mmbscatter$).  The library also utilizes the phenomenological binary evolution model and varies the SMBH binary lifetime, $\tlifetime$, and the hardening power-law index at small separations $\hardnuinner$.

The differences in GWB spectra for systematic variations in \libphenom{} model parameters are shown in Figure~\ref{fig:gwb_anatomy}.  The overall amplitude of the GWB spectrum varies most significantly in the left and top-middle panels, indicating that the GWB amplitude is most sensitive to parameters determining SMBH masses ($\gsmfmass$, $\mmbamp$) and the SMBH binary number density ($\gsmfnorm$).  In the lower-middle panel, we see that increasing scatter in the \mmbulge{} relationship ($\mmbscatter$) also increases the GWB amplitude. This owes to the fact that larger scatter increases the effective SMBH masses through Eddington bias: because low-mass SMBHs are more numerous, their scatter towards higher masses outnumbers the scatter of the rarer, higher-mass SMBHs towards lower values.  Notice that variations in SMBH mass, GSMF turnover mass, and \mmbulge{} scatter (parameterized by $\mmbamp$, $\gsmfmass$, and $\mmbscatter$, respectively) all produce qualitatively similar changes in the GWB spectra. Higher masses preferentially increase the low-frequency amplitudes, thereby steepening the spectra at higher frequencies ($f \gtrsim 1 \pyr$). This occurs as rare high-mass binaries contribute less to the GWB at higher frequencies, due to their rapid evolution at smaller separations (see \S~\ref{sec:meth_binary_pops}).  The $\mmbscatter{}$ parameter shows this frequency-dependent effect even more prominently, as it preferentially affects the highest SMBH mass bins where the gradient in SMBH number density with respect to mass is steepest.

The shape of the spectrum at low frequencies is determined by the binary hardening rate $da/dt$ (as introduced in \S~\ref{sec:meth_binary_evo}), which includes the interaction of binaries with their nuclear galactic environments. Recall that in our models, the binary lifetime is an input parameter, one that is varied in the top-right panel of Figure~\ref{fig:gwb_anatomy} and kept fixed at $\tau_f = 1.0$ Gyr in all other panels. Consequently, for a given $\tau_f$, binaries of different masses enter the GW regime at different frequencies. Some of the variations in low-frequency spectral shape seen when the mass-determining parameters ($\mmbamp$, $\mmbscatter$, \& $\gsmfmass$) are varied and hardening parameters are kept fixed can therefore be attributed to $\tau_f$ being the same for all binary masses.  For our models with fixed total binary lifetimes, populations with lower masses tend to have stronger low-frequency turnovers as lower-mass binaries enter the GW regime at higher frequencies.  Equivalently, lower mass systems spend more time at higher frequencies, meaning that their environmentally driven evolution must have proceeded even faster at lower frequencies.

While there is some degeneracy across all parameters, only the two parameters that directly affect the average binary mass ($\gsmfmass$ and $\mmbamp$) produce mostly degenerate spectral changes.  As mentioned above, even the \mmbulge{} scatter parameter ($\mmbscatter$), which also changes the average binary mass, is noticeably distinct.  This speaks to the possibility of independently constraining multiple parameters with a sufficiently high signal-to-noise ratio, even without appealing to additional information content such as sky anisotropy, individual continuous-wave sources, or electromagnetic counterparts and other multi-messenger constraints.

\begin{figure}
    \centering
    \includegraphics[width=1.0\columnwidth]{{{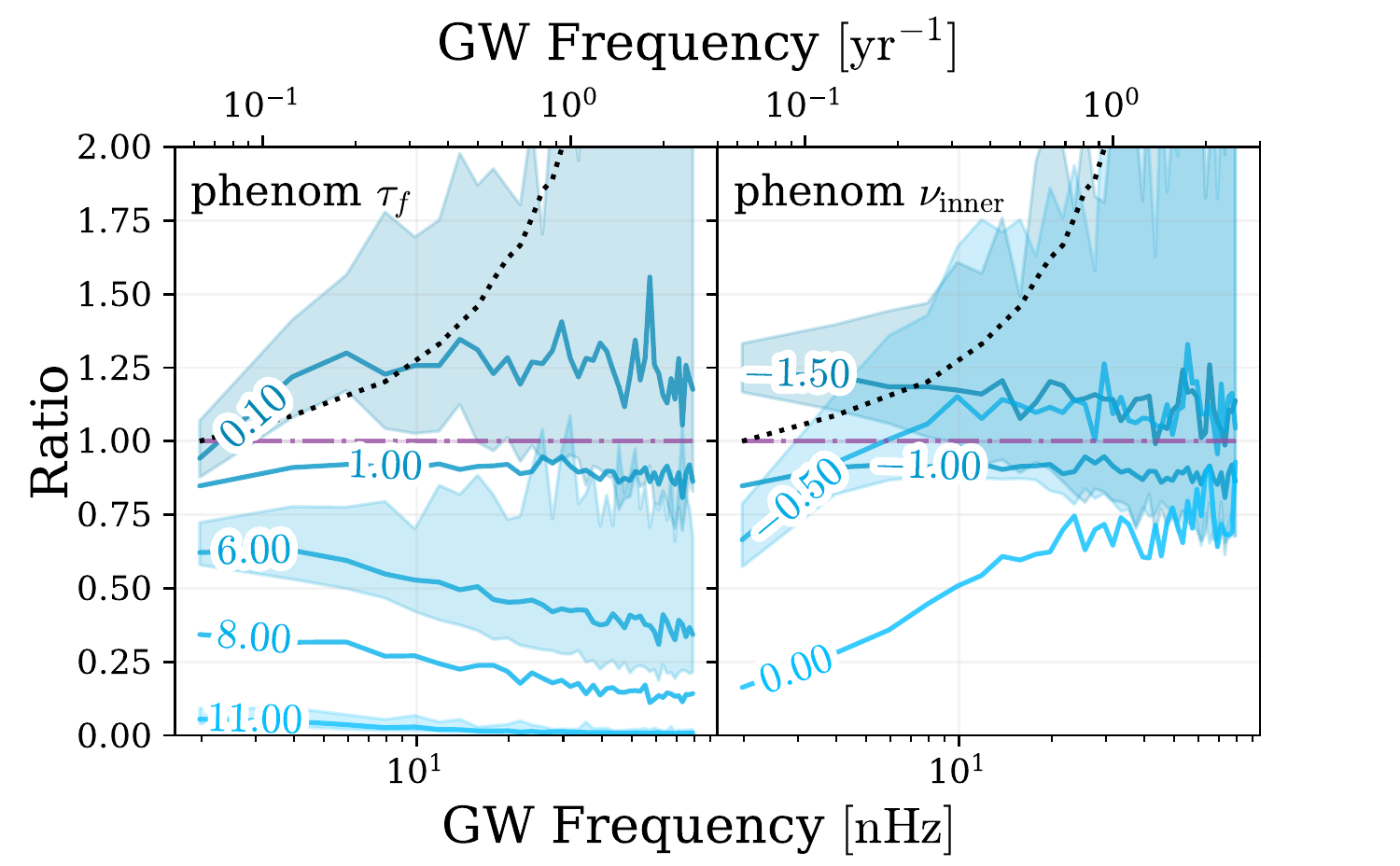}}}
    \caption{Comparison of GWB spectra for systematic variations in two of the six model parameters ($\tlifetime$ and $\hardnuinner$) normalized to a model in which binaries evolve only because of GW emission.  Colors and line styles are as in Fig.\ \ref{fig:gwb_anatomy}.  The phenomenological hardening models show substantial deviations from the GW-only hardening models, especially for parameters far from our fiducial values of $\tlifetime = 1\ \mathrm{Gyr}$ and $\hardnuinner = -1$.  In particular, the spectra are suppressed at the lowest frequencies for larger values of $\hardnuinner$ and produce a more pronounced turnover.}
    \label{fig:gwb_anatomy_gwonlycomp}
\end{figure}

As introduced in \S~\ref{sec:meth_binary_evo}, in addition to \libphenom{}, we also use a library with the same variations in GSMF and \mmbulge{} parameters, but using GW-only evolution instead of the phenomenological model.  We refer to this four-dimensional parameter space $\{\gsmfnorm, \gsmfmass, \mmbamp, \mmbscatter\}$ as \libgwonly{}.  Figure~\ref{fig:gwb_anatomy_gwonlycomp} shows a comparison of GWB spectra with variations in the \libphenom{} model parameters versus spectra from the \libgwonly{} models.  For longer binary lifetimes ($\gg \tr{Gyr}$), fewer systems are able to coalesce, and the GWB amplitude is noticeably diminished at all frequencies.  For shorter lifetimes ($\lesssim \tr{Gyr}$), non-GW hardening is still important at low frequencies within the 15 yr NANOGrav band.  This leads to binaries evolving faster than the GW-only prediction, fewer binaries existing at these frequencies, and thus attenuated GW emission producing a low-frequency turnover \citep{Kocsis+Sesana-2011, Ravi_2014}.  Moderate inspiral times ($\sim \tr{Gyr}$) produce the closest match between the phenomenological and GW-only models, but still show a slight turnover in addition to an amplitude $\sim 10\%$ lower at all frequencies.


\subsection{Interpolation of Population Synthesis Models with Gaussian Processes}
\label{sec:interpolation}

In order to infer the properties of SMBH binary populations that are consistent with the GWB, we need to compare the theoretically expected GWB spectra from \holodeck{} with the observed NANOGrav data. Previous work used analytic expressions for this, e.g., by fitting the GWB spectra with a single power law (for a population of circular binaries purely driven by GWs) or a broken power law (to capture the turnover produced by environmental interactions; \citealt{2015PhRvD..91h4055S}). However, the properties of the SMBH binary population are only indirectly extracted from these fits, and disentangling potential covariances between population parameters is challenging. To overcome this limitation, \citet{2017PhRvL.118r1102T} developed a modeling framework that directly links the properties of the GWB spectrum to the binary population parameters by training Gaussian processes (GPs) on simulated GWB spectra from population-synthesis models. Here we adopt this approach to interpolate the strain of the GWB across simulated \holodeck{} libraries, generated in discrete points of the binary parameter space, to accurately predict the GWB spectrum at any point in the space.

GPs provide a powerful interpolation method that parameterizes noisy data in terms of a multivariate Gaussian distribution with a mean vector and covariance function \citep[see][for a review]{2022arXiv220908940A}. The covariance functions can be custom built from a suite of versatile kernel functions allowing for quick adaptability to a variety of complex parameter spaces. While GPs are not sparse and lose efficiency in high-dimensional spaces (e.g., greater than a few dozen), one key advantage of GP regression is that it provides an estimate of the uncertainty in the interpolation process (i.e., the prediction is probabilistic). Importantly, one can use this in an iterative process to adapt and improve the fitting. Additionally, the GP uncertainty can be propagated forward to our final statistics, allowing for a full marginalization over the interpolation uncertainties.

The GPs are trained on \holodeck{} GWB spectra using the \texttt{George} GP regression library \citep{2015ITPAM..38..252A}, as in \citet{2017PhRvL.118r1102T} and \citet{NANOGrav11yrGWB}. To capture fluctuations that arise directly from the discrete nature of the binary population, we train the GPs at each sampling frequency of the GW spectrum, $f_i$. Since GP regression assumes that the interpolated quantity (here the strain of the GWB) is smooth with respect to the interpolation variables (here the model parameters).  The use of an independent GP at each frequency thus enforces smoothness in the GW spectrum across model parameter space at a given frequency, but not across frequencies.  Because the binary population is independent at each frequency, smoothness across frequencies is not expected in general.  Two separate GPs are trained per frequency, one on the median values of $\log_{10}\left(h_c^2(f_i)\right)$, and one on its standard deviation.  This allows us to predict both the typical value and the typical spread of the strain, and to account for the uncertainty in each value's interpolation separately.

We select the training set (i.e., the library generation points that make up our model grid) from our multi-dimensional parameter space using Latin hypercube (LHC) sampling \citep[e.g., see][and references therein]{2018PhRvD..98h3017T}.\footnote{One-dimensional LHC sampling divides the cumulative density function into a number of equal partitions, and then chooses a random data point in each partition.  Sample points in multiple dimensions are randomly combined.  This approach ensures coverage of the domain, similar to a uniform grid, while not wasting samples at identically placed grid edges.}  This offers an efficient method to generate a near-random set of parameter values, representative of the entire parameter space, with relatively few points. Since we aim to explore high-dimensional spaces, this type of sampling is necessary to keep the total number of simulations computationally tractable.

The training of the GPs proceeds as follows. Using the LHC method, we draw $s$ samples in the binary parameter space. For each sample, we produce $r$ realizations of the GWB spectrum using \holodeck{} and calculate the median and standard deviation of $\log_{10}(h_c^2)$ at each frequency $f_i$. These means and standard deviations constitute the inputs for the training of the two GPs.  For each point in the training set, GPs require the value of the quantity on which they are trained (here the median or standard deviation of $\log_{10}[h_c^2(f_i)]$) and optionally its uncertainty.  Including uncertainties on the input values helps to avoid over-fitting.  We adopt the standard uncertainties for sample mean and sample standard deviation. When training on the median, we estimate the uncertainty as the standard deviation divided by ${r}^{1/2}$; and for training on the standard deviation, the uncertainty is given by the standard deviation divided by $[2(r-1)]^{1/2}$.

\begin{figure}
    \centering
    \includegraphics[width=1.0\columnwidth]{{{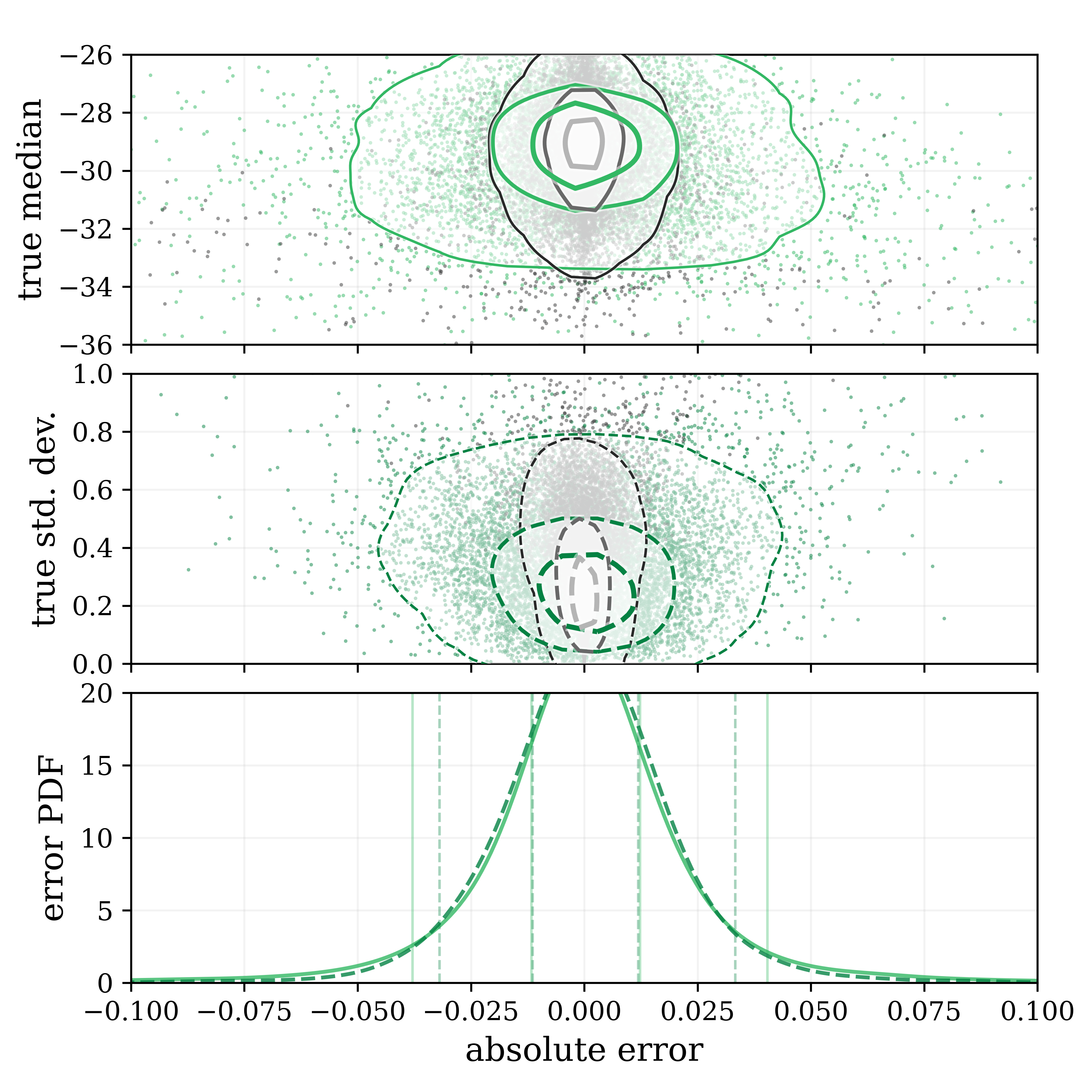}}}
    \caption{Accuracy of trained Gaussian process (GP) interpolants for our fiducial, six-dimensional parameter space \libphenom{}, aggregated over the first $5$ frequency bins used in our primary analyses. The \textit{top} panel shows the error ($\equiv \tr{predicted} - \tr{true}$) on the median $\log_{10}\hc^2|_\tr{predicted}$ (solid lines) for the training set (gray) and validation set (green), while the \textit{middle} panel shows the error on the standard deviation (dashed lines).  Contours contain $20\%$, $50\%$, and $90\%$ of the population.  The \textit{bottom} panel shows the distributions of errors for both validation sets. The vertical lines bound the $50\%$ and $90\%$ region of the errors. The GPs predict 99.4\% (98.5\%) of the validation set for medians (standard deviations) within 10\% of the actual value.  Both the training and validation sets contain 2,000 sample points, with medians and standard deviations calculated over 2,000 realizations, which we have found to be more than sufficient in accurately training the GPs.}
    \label{fig:gp_errors}
\end{figure}

To test the performance of the GPs, we create a validation set with points in the SMBH binary parameter space that were not included in the training set. For each validation point, we calculate the median and standard deviation of $\log_{10}(h_c^2)$, both with GPs and with \holodeck{} simulations. For comparison purposes, we label the value obtained from GPs as the ``predicted'' value, while the \holodeck{} values are considered to be the ``true'' value. Based on this, an error (i.e., predicted minus true) can be calculated for the GP interpolation performance.

We used this approach to test a variety of kernels (i.e., covariance functions) along different directions in parameter space to determine which combination most accurately captured each parameter's response to changing the GWB. We determined that two types of kernels were necessary: rational quadratic kernels for the phenomenological timescale $\tlifetime$ and the hardening power-law index $\hardnuinner$, and squared exponential kernels for the remaining parameters. An iterative process of checking the performance of the GPs was used to determine the necessary number of LHC sample points and \holodeck{} realizations to converge on a sufficient accuracy level.  The performance of the GP trained on the median values is more sensitive to the choice of the number of sample points, while the performance of the GP trained on the standard deviations is more sensitive to the choice of the number of realizations. We found that training on $s=$ 2,000 LHC samples with $r=$ 2,000 \holodeck{} realizations at each sample point was more than enough to acquire the desired accuracy level. Figure \ref{fig:gp_errors} shows the response of GPs trained on the \libphenom{} library for the $5$ frequencies used in our analysis.  The reconstruction is quite accurate, with 99.4\% (98.5\%) of the test set cases for medians (standard deviations) falling within 10\% of the actual value---significantly smaller than the standard deviation across spectra realizations.


\subsection{Fitting Simulated GWB Spectra to PTA Observations}
\label{sec:meth_ceffyl}

In \citet{NANOGrav11yrGWB}, once GPs were trained, they were inserted into the full PTA likelihood calculation in order to obtain posteriors on the SMBH binary population parameters. As PTA data sets have grown in size and with the new discovery of HD correlations, the likelihood computation time has increased. As such, inserting two GPs into the 15 yr data set's likelihood calculation in order to obtain posteriors for \libphenom{} was not an efficient analysis approach.

Instead, we use the \texttt{ceffyl} package \citep{Lamb+2023} to fit the interpolated GWB spectra to the previously computed free-spectrum posteriors of the cross-correlated timing-residual PSD. Fitting on intermediate PTA analysis products, such as the free-spectrum posteriors, offers a substantial speed-up by factors of $10^{2}-10^{4}$ compared to directly fitting the full likelihood of timing residuals. Importantly, the resulting posterior distributions of GW spectral model parameters achieved by \texttt{ceffyl} have been found to be nearly identical to those obtained from the full likelihood approach.

In detail, we expand the likelihood, $\mathcal{L}\left(\vec{d}|\vec{\Theta}\right)$, where $\vec{d}$ is the PTA data (e.g., the TOAs) and $\vec{\Theta}$ are the SMBH binary population parameters (e.g., the parameters from \libphenom{}), by inserting an intermediate data product such as the free-spectrum posteriors ($\log_{10}(\rho_i)$). Then, instead of directly calculating the fit of a GWB spectrum (generated by the trained GPs for a given draw of SMBH binary population parameters) to the TOAs, we compute the probability that a given GWB spectrum is supported by the free-spectrum posteriors. The expanded likelihood function is now given by
\begin{align} \label{eq:ceffyl}
    &\mathcal{L}\left(\vec{d}|\vec{\Theta}\right) \propto \nonumber \\
    &\prod_{i=1}^{N_f} \int \mathrm{d}\left(\log_{10}\rho_i\right)\  p\left(\log_{10}\rho_i|\vec{d}\right) p\left(\log_{10}\rho_i|\vec{\Theta}\right),
\end{align}
where $N_f$ is the number of Fourier components used in the GWB analysis (5 or 14, but see also \S~\ref{sec:data} and \citetalias{nanograv_15yr_gwb}), $p\left(\log_{10}\rho_i|\vec{d}\right)$ is the posterior probability density of $\log_{10}(\rho_i)$ (i.e., the free-spectral posteriors) which are represented by highly optimized kernel density estimators, while $p\left(\log_{10}\rho_i|\vec{\Theta}\right)$ is the probability of $\log_{10}\rho_i$ given a GWB spectrum from the trained GPs. Since the GPs are trained on the median and standard deviation of the characteristic strain $\log_{10}(h_c^2)$, these provide the mean and variance of a Gaussian when calculating $p\left(\log_{10}\rho_i|\vec{\Theta}\right)$. The above likelihood is sampled through MCMC techniques to obtain the resultant posteriors on $\vec{\Theta}$.

\begin{figure}
    \includegraphics[width=1\columnwidth]{{{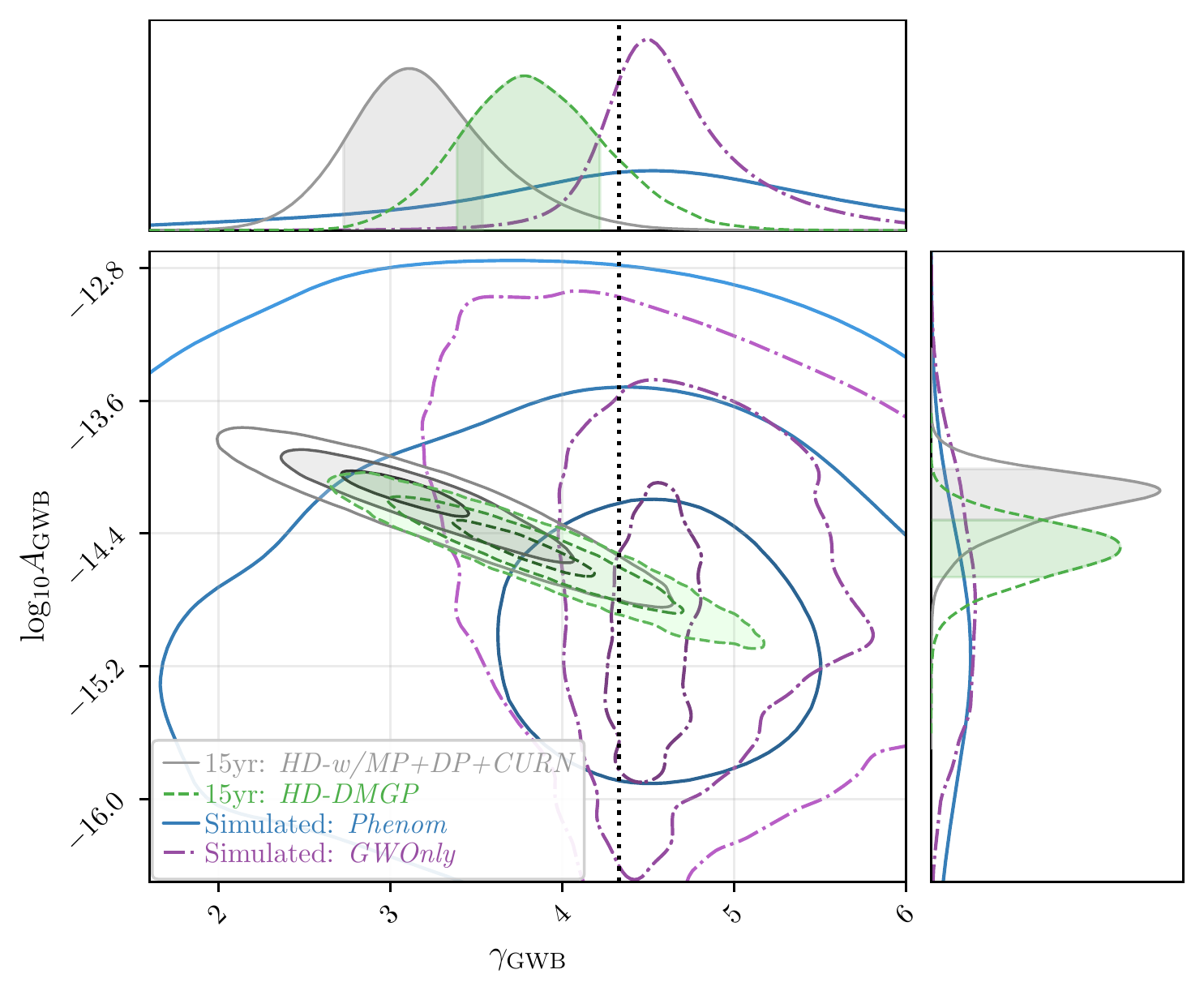}}}
    \caption{Power-law amplitude ($A$) and spectral index ($\gamma$) from purely power-law fits to HD free-spectrum model posteriors from the 15 yr data set, compared to simulated GWB spectra from \holodeck{} libraries.  Data set fits include both the 15 yr \hdall{} and \hdgp{} models for comparison.  Fits to the lowest 5-frequency bins of spectra from two \holodeck{} libraries are shown: the self-consistent phenomenological binary evolution model (\libphenom{}) and the purely GW-driven evolution model (\libgwonly{}). We show 1, 2, \& 3$\sigma$ contours for each.  The analytic, GWB PSD power-law index of $\gamma = 13/3$ is shown as reference (black, dashed).  The spectral shape of the HD signal present in the 15 yr data set is broadly consistent with expectations for a GWB from binary SMBH population. The amplitude is towards the higher end of predictions and the recovered spectral index deviates from the idealized power law in similar ways as the phenomenological binary evolution model.
    }
    \label{fig:amp_index}
\end{figure}

While all of the libraries generated for GP training draw uniformly from the SMBH binary population parameter space, when we perform the MCMC analysis, we have the opportunity to place different priors onto each parameter. For the analysis in this paper, we utilize two distinct prior set-ups: a uniform prior and a set of astrophysical priors based on galaxy observations (e.g., see Table~\ref{Table:sam_params}). When relevant, we denote the prior distribution shape in combination with the library designation as e.g., \libphenomuniform{} or \libphenomastro{}   (see Table~\ref{Table:models}).


\section{Results}
\label{sec:results}
We simulate populations of SMBH binaries using a phenomenological (\libphenom{}) and GW-only (\libgwonly{}) model. We create \holodeck{} libraries of GWB spectra at fixed points of the SMBH binary parameter space and interpolate them with GPs. We fit the models to the 15 yr free-spectrum posteriors considering the \hdall{} as the fiducial 15 yr NANOGrav results for this analysis (but we also fit the \hdgp{} posteriors for comparison) using both uniform and astrophysically motivated priors (see Table~\ref{Table:sam_params}). As shown in Table~\ref{Table:models}, the \libphenom{} library is fit against the data using both uniform priors and astrophysically informed priors (\libphenomuniform{} and \libphenomastro{}), while the \libgwonly{} library is fit only with uniform priors (\libgwonlyuniform{}). Our results are summarized as follows.

In all of our analysis, we find that the NANOGrav 15 yr data set is consistent with a GWB produced by a population of SMBH binaries.  In the first, most simplified approach, power-law fits\footnote{Note that NANOGrav constraints are derived primarily at lower frequencies.  Fitting power laws, and extrapolating the amplitudes to $f=\lr{1 \, \yr}^{-1}$ can lead to amplitudes that differ more significantly at this frequency than at $f=\lr{10 \, \yr}^{-1}$, for example.  See Appendix~\ref{sec:app_lit_models}.} to both the observed GWB spectrum and those from simulations produce amplitudes and spectral indices that overlap in the 2- and 3-$\sigma$ regions depending on model (\S~\ref{sec:results_gwb_plaw}).  The remainder of this section presents the results of our systematic approach of fitting simulated SMBH binary populations to the data, which yield more realistic GWB spectra that match the 15 yr results (\S~\ref{sec:results_binary_gwb}). From these fits we obtain posterior distributions on uncertain astrophysical parameters of the SMBH population synthesis models (\S~\ref{sec:results_astro_posteriors}) and make predictions from our models for the properties of the population of SMBH binaries that produce the GW observations (\S~\ref{sec:results_astro_constraints}).

\begin{figure*}
    \includegraphics[width=1\textwidth]{{{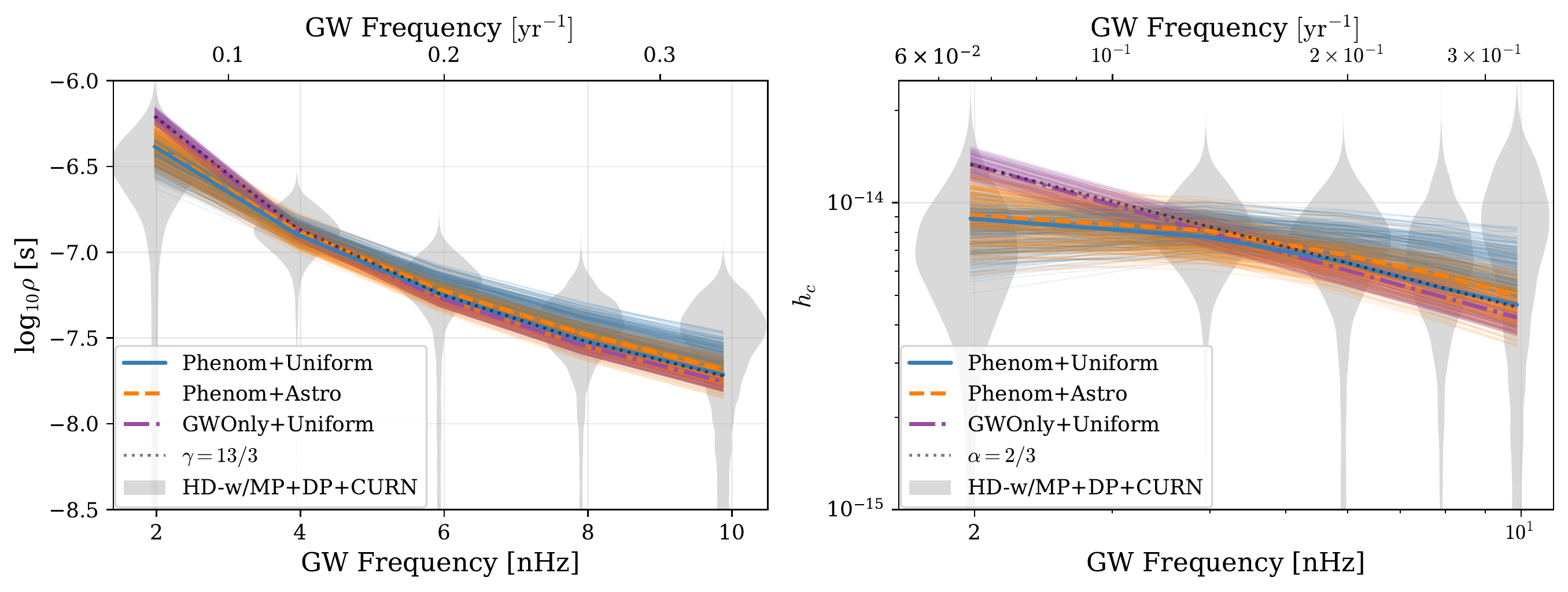}}}
    \caption{GWB spectra from simulated SMBH binary populations that best fit the 15 yr free-spectrum data. The left panel shows the square-root of cross-correlated timing-residual power ($\rho$) and the right panel shows characteristic strain ($h_c$). GP interpolated spectra are shown, with thick lines showing maximum likelihoods and thin lines showing $200$ random draws from the posteriors. The \libphenom{} library uses self-consistent binary evolution models, while \libgwonly{} assumes purely GW-driven evolution.  For the former, fits using uniform priors (\libphenomuniform{}) are compared against more informed, astrophysically motivated ones (\libphenomastro{}).  Power-law fits to the 15 yr spectra with $A_\mathrm{yr} = 2.1 \times 10^{-15}$ are also shown as dotted black lines.}
    \label{fig:free_spec_fit}
\end{figure*}

\subsection{Comparison of idealized power-law fits to GWB spectra}
\label{sec:results_gwb_plaw}

The approach of fitting simple power-law models to the GWB is a common one in the literature. While idealized power-law fits to GWB spectra neglect most of the information imprinted by astrophysical processes on the background, they are effective in broadly examining the consistency between simulated binary populations and PTA data sets. Therefore, we carry out this straightforward analysis as a first check of the \libphenom{} and \libgwonly{} libraries before implementing the full methodology described in \S~\ref{sec:interpolation} - \ref{sec:meth_ceffyl}. In practice, we constrain the amplitude, $A$, and slope, $\gamma$, of an idealized power-law GWB spectrum (in timing-residual PSD; Equation~\ref{eq:psd_plaw}) with a non-linear least-squares fits to the GWB spectra from each realization of the binary population from the \libphenom{} and \libgwonly{} libraries using the five lowest-frequency bins. We then compare these to the results of power-law fits of the 15 yr data, illustrating their overlap in the $A-\gamma$ parameter space.

In Figure~\ref{fig:amp_index}, we show the range of GWB amplitudes and spectral indices (in timing-residual PSD) based on these fits.  We see that the amplitudes and power-law indices vary significantly across the simulated GWB spectra.  Even for the \libgwonly{} models, which match the premise of the analytic $\gamma = 13/3$ ($\alpha=2/3$) models, our simulations yield indices typically varying from $4$--$5.5$ in the 95\% credible region. Recall that even GW-only binary evolution with circular orbits does not produce a pure power-law spectrum, owing to the steepening of the spectrum at higher frequencies where the finite number of binaries in each frequency bin becomes important. The slight offset of the \libgwonly{} models towards steeper values of $\gamma > 13/3$ reflects this higher-frequency spectral steepening caused by finite-number effects.  The \libphenom{} libraries, which self-consistently model the effects of environmental interactions on binary evolution, produce much wider ranges of spectral indices as disparate as $2$--$7$, with lower values corresponding to shallower characteristic strain spectra (i.e., increasing across the lowest five frequency bins). Note that we exclude from Figure~\ref{fig:amp_index} a small number ($\sim 1\%$) of \libphenom{} samples in which all binaries stall and thus produce zero GWB.

We also show the power-law parameter posteriors for the fiducial \hdall{} free-spectrum posteriors, and the \hdgp{}, which we use for comparison.  While the 15 yr free-spectrum posteriors are not perfectly fit by power-law models (see Figure~\ref{fig:15yr_spectrum}), the differences between these two models highlight that the measurement of a spectral index is particularly sensitive to choices of fit in the 15 yr data, and to features in particular frequency bins \citepalias{nanograv_15yr_gwb}.

We use the $A-\gamma$ fits as a general measure of parameter space coverage.  Figure~\ref{fig:amp_index} demonstrates that the range of simulated populations is able to reproduce the measured GWB within the two-sigma curve of the \libphenom{} library, and between the two- and three- sigma curves of the \libgwonly{} library.

\begin{figure*}[tbh!]
    \centering
    \includegraphics[width=\textwidth]{{{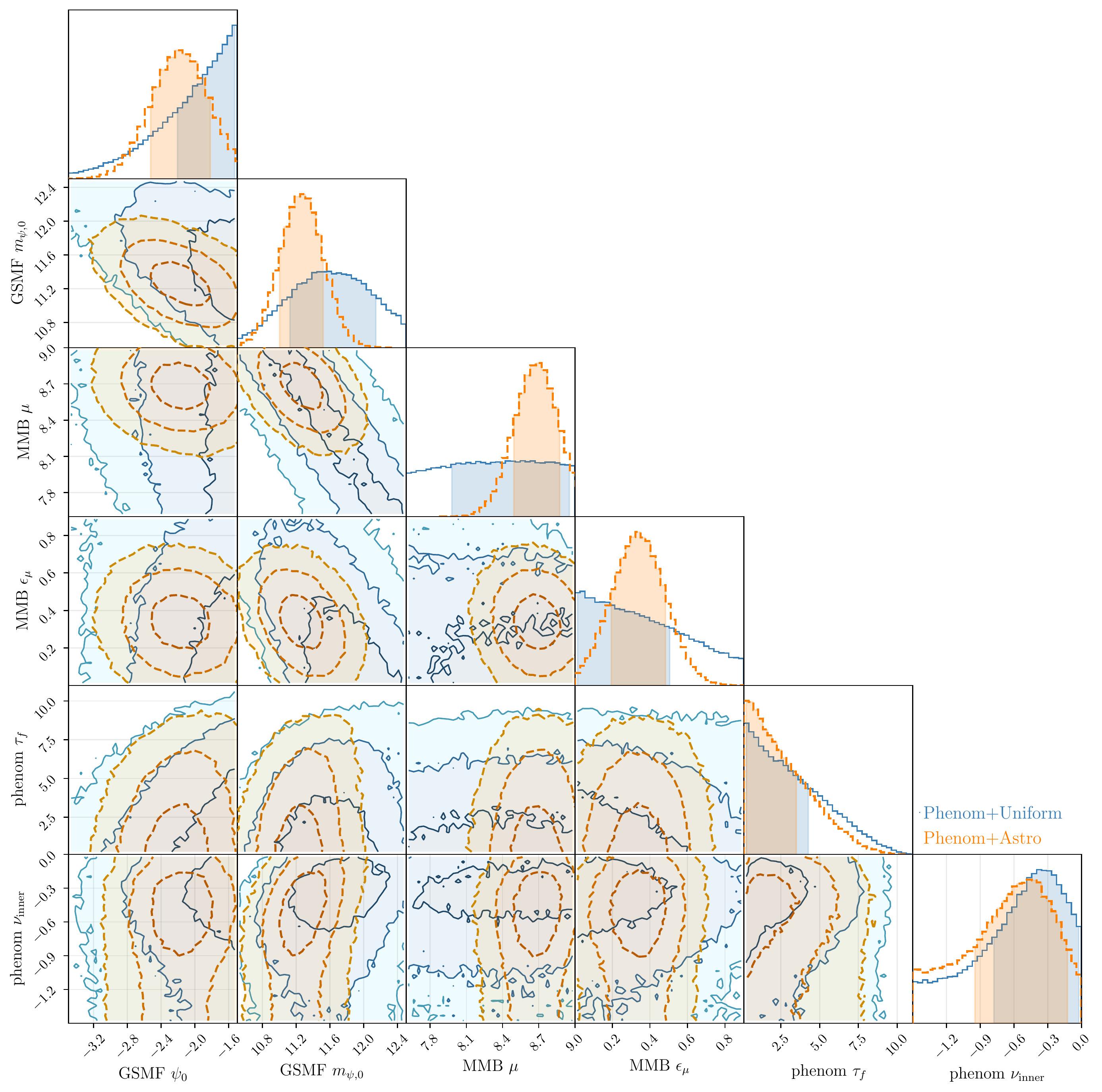}}}
    \caption{Binary evolution parameter posteriors from fitting against the 15 yr \hdall\ free spectrum.  The parameters correspond to the phenomenological library \libphenom{}: binary lifetime ($\tlifetime$), the hardening power-law index ($\hardnuinner$), the GSMF normalization ($\gsmfnorm$) and characteristic mass ($\gsmfmass$), and the \mmbulge{} normalization ($\mmbamp$) and scatter ($\mmbscatter$). The fiducial \holodeck{} spectra library is fit to the lowest five frequencies utilizing two separate priors. \libphenomuniform{} uses flat priors across all parameter spaces (blue, solid-line contours), while \libphenomastro{} (orange, dashed-line contours) uses constrained priors on the GSMF and \mmbulge{} parameters that encapsulate observational measurements for these parameters. The contours in two dimensions correspond to the 1, 2, and 3$\sigma$ regions, and the shaded regions in the one-dimensional plots are 1$\sigma$ regions. Although individual parameters are only weakly constrained, the data strongly prefer efficient mergers in high-mass systems.}
    \label{fig:posteriors}
\end{figure*}


\subsection{The GWB is Consistent with Expectations from Populations of SMBH Binaries}
\label{sec:results_binary_gwb}

The consistency between the 15 yr NANOGrav data set and GWBs produced by SMBH binaries is best supported by an analysis of the full range of astrophysical information contained in the free-spectrum posteriors. Fitting the GPs trained on the \libgwonly{} and \libphenom{} libraries to the 15 yr data (with uniform and astrophysically motivated priors) facilitates a comparison of observations to the GWB spectra from SMBH binary populations that is agnostic to any particular spectral model (including a power law).

Figure~\ref{fig:free_spec_fit} shows GWB spectra produced by our simulated SMBH binary populations that accurately fit the 15 yr \hdall\ free spectrum.  As mentioned above, the \libphenom{} library is fit both with uniform  and astrophysically informed priors (\libphenomuniform{} and \libphenomastro{}), while the \libgwonly{} library is fit only with uniform priors (\libgwonlyuniform{}).  Thin curves show 200 random draws of the binary parameter posterior distributions for each of the above models, with thick lines denoting the maximum likelihood spectra for each model.

Both libraries are able to fit the GWB within the 15 yr posteriors.  However, the \libgwonly{} spectra have more difficulty matching the data, as indicated by their preference for the edges of the 15 yr free-spectrum posteriors in the highest and lowest frequency bins, and the best fit spectrum missing the highest probability regions of the 15 yr GWB data.  As a comparison, power-law fits are shown for the idealized $\gamma=13/3$ ($\alpha = 2/3$) spectral indices obtained from analytic calculations of SMBH binaries \citep{Phinney-2001}.  The \libgwonly{} models, which more closely resemble these analytic estimates, tend to be steeper than the bulk of the 15 yr distributions.  In contrast, the maximum-likelihood spectra and likelihood draws from the \libphenom{} model exhibit noticeable spectral turnovers to match the 15 yr data.  While these results are suggestive of a low-frequency turnover or flattened spectrum, they are still consistent with an $\alpha = 2/3$ power law and the associated GW-driven evolution.


\subsection{Parametric Constraints on SMBH Binary Models}
\label{sec:results_astro_posteriors}

The MCMC exploration of the likelihood in Equation~(\ref{eq:ceffyl}) returns constraints (posterior distributions) on the parameters of the population synthesis models based on the observed GWB spectrum. The peaks of the marginalized posteriors indicate the most likely values of the parameter space that the binary population must occupy in order to produce the 15 yr free-spectrum posteriors.  Figure~\ref{fig:posteriors} shows the posteriors of these binary population parameters for the \libphenom{} binary evolution model.  Results are compared for different prior choices, i.e. the \libphenomuniform{} and the \libphenomastro{} fits.  Owing to the substantial uncertainty in the GWB spectrum at NANOGrav's current sensitivity, the posteriors are sensitive to the assumed priors, and only weak parameter constraints can be made.

However, we can identify some general trends among the preferred parameter values.  The measured amplitude of the GWB strongly prefers a combination of efficient mergers occurring in high-mass systems. The data favor short binary lifetimes ($\tlifetime$), high GSMF number densities ($\gsmfnorm$), and high characteristic masses ($\gsmfmass$, $\mmbamp$). It is worth noting that the range of priors in the \libphenomuniform{} fits is quite wide compared to typical values adopted in the astronomical literature (see references with Table~\ref{Table:sam_params}).  While the parameters in the \libphenomastro{} model are more constrained, the posteriors are still fairly broad.  Because our models utilize simplified analytic prescriptions for each physical component, we use broader parameter distributions in \libphenomuniform{} in part to introduce some added flexibility.  None the less, Figure~\ref{fig:posteriors} demonstrates that the posteriors almost uniformly favor parameters that produce larger GWB amplitudes (e.g., see also Figure~\ref{fig:gwb_anatomy}).  This suggests that the amplitude of the GWB inferred from the 15 yr data set is difficult to reach with standard values of some astrophysical parameters.

Very long binary lifetimes are disfavored for both \libphenomuniform{} and \libphenomastro{}. A large fraction of binaries with such long lifetimes would fail to reach the NANOGrav frequency band, resulting in lower GWB amplitudes, inconsistent with the GW data. Flatter values of the hardening rate power-law index, $\hardnuinner \gtrsim -1.0$, are also preferred, as they produce spectral turnovers in the lower frequency bins (see Figure~\ref{fig:binary-evo}) resembling what is seen in the 15 yr data.  Steeper values of $\hardnuinner$ correspond to binaries that spend more time at $\sim$10$^{-2}$ -- 1 pc separations and transition into the GW-dominated regime earlier, at frequencies below the PTA band.  Flatter values of $\hardnuinner$ correspond to very efficient inspiral through this range of separations, leading to environmentally driven evolution even in the lower PTA band which produces noticeable GWB attenuation and a sharp low-frequency spectral turnover (see Figure \ref{fig:gwb_anatomy_gwonlycomp}).  We note that this is dependent on the parameterization of our binary evolution model and the parameters varied in the \libphenom{} library.  Steeper evolution profiles at very large separations ($\sim$kpc), i.e., larger values of $\hardnuouter$ could similarly produce low-frequency turnovers (but this is not explored in this paper, since $\hardnuouter$ is kept fixed throughout).  In either case, efficient binary inspiral in the environmentally driven regime produces the most noticeable spectral turnovers.

The posteriors for GSMF and \mmbulge{} parameters differ noticeably for uniform versus astrophysical priors, unlike the binary inspiral parameters.  The posteriors for the normalization and the characteristic mass of the GSMF ($\gsmfnorm$ and $\gsmfmass$) favor values at the higher end of the prior range, especially in the $\libphenomuniform{}$ fits, in which the galaxy number densities are pushed against the edges of the prior.
We also see higher values of these posteriors when binary lifetimes are longer, such that larger fractions of binaries stall before reaching the PTA band.

Unsurprisingly, the GSMF characteristic mass ($\gsmfmass$) is almost entirely degenerate with the \mmbulge{} mass normalization ($\mmbamp$), as indicated by the diagonal band in the respective 2D posterior.  The scatter in the \mmbulge{} relationship ($\mmbscatter$), however, shows different trends.  This is likely due to two factors.  First, increasing $\mmbscatter$ primarily increases the GWB amplitude in the lower frequency bins (Figure~\ref{fig:gwb_anatomy}), as larger scatter preferentially increases SMBH masses and higher masses are more prevalent at lower frequencies (discussed more below).  Secondly, larger values of $\mmbscatter$ also produce significant variance across multiple population realizations which may decrease the aggregated likelihoods when calculating fits.

\begin{figure*}[tbh!]
    \centering
    \includegraphics[width=\textwidth]{{{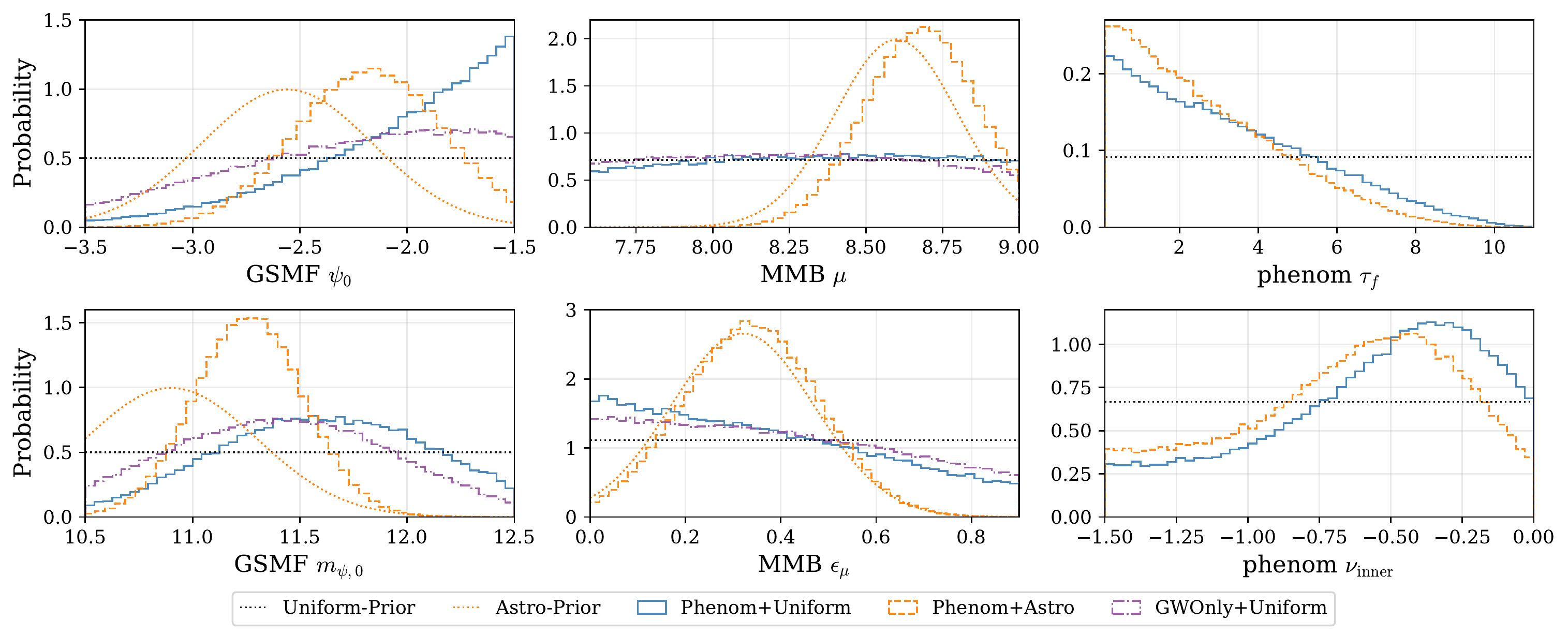}}}
    \caption{One-dimensional binary evolution parameter posteriors from fitting against the 15 yr \hdall{} free spectrum.  Priors (dotted) and posteriors (solid) are shown for the \libphenom{} model with both uniform and astrophysical priors. Note that the astrophysical and uniform priors are identical for $\tlifetime$ and $\hardnuinner$ (see Table~\ref{Table:sam_params}). The \mmbulge{} parameters are the least constrained (i.e., the posteriors largely resemble the priors), while the GSMF parameters are the most constrained. This is consistent with our expectations given each parameter's impact on the simulated GWB (see Figure \ref{fig:gwb_anatomy}). However, these 1D projections can miss significant 2D constraints due to various projection effects (e.g., the combined \mmbulge{} mass normalization and $\gsmfmass$ 2D space is much more constrained than each parameter individually, as seen in Figure \ref{fig:posteriors}). Overall, there is broad consistency between the posteriors for the \libphenom{} models, regardless of which prior is used, showing that these constraints are coming directly from the GWB spectrum itself and are not based solely on our binary population model construction.  Distributions are also shown for the \libgwonly{} model, which gives posteriors broadly consistent with the phenomenological models. However, they are peaked at slightly lower parameter values, indicating an overall lower amplitude for their simulated GWB. This is most likely due to this model's rigidity, which does not allow for significant deviations from a power-law-like GWB and leads to an inability to capture the behavior of the lowest frequency bin in the 15 yr free spectrum (see Figure \ref{fig:free_spec_fit}). }
    \label{fig:1Dhist_posteriors}
\end{figure*}

In Figure~\ref{fig:1Dhist_posteriors}, we compare the one-dimensional distributions of parameter priors versus posteriors for the \libphenomuniform{}, \libphenomastro{}, and \libgwonlyuniform{} models.  Particularly in the case of the \mmbulge{} parameters, we see that the posteriors closely follow the priors.  While still consistent with the priors, the GSMF parameters are pushed noticeably towards higher values, even for the \libphenomastro{} fits.

Figure~\ref{fig:1Dhist_posteriors} also shows fits using the \libgwonly{} library.  Note that this library does not include the $\tlifetime$ or $\hardnuinner$ parameters by definition.  The posterior distributions for the GSMF and \mmbulge{} parameters are generally consistent in both the \libphenom{} and \libgwonly{} models, with only weak constraints.  While \libgwonly{} shows the same preference for high values of $\gsmfnorm$ as the phenomenological model, the preference is less pronounced.  This is likely due to the decrease in GWB power at the lowest frequencies when a spectral turnover is induced, which is consistent with the covariance seen in Figure~\ref{fig:posteriors} between binary lifetime and GSMF normalization.

It is important to note that these parametric constraints must be interpreted in the context of the semi-analytic binary evolution models used to generate the binary populations and corresponding GWB spectra.  For example, the usage of a fixed-time phenomenological binary evolution model is forcing a particular relationship between typical binary masses and the degree of low-frequency spectral turnover.  Another model, in which the degree of environmental coupling scales differently with binary mass \citep[or similarly, host-galaxy properties; e.g.,][]{Kelley+2017a}, may produce different dependencies and thus different posteriors.  We are also assuming a fixed \mmbulge{} relationship for all redshifts, while the canonical \mmbulge{} relationship in the literature is specifically calibrated to the local Universe.  Our values of $\mmbamp$ and $\mmbscatter$ for both the \libphenom{} and \libgwonly{} models should thus be interpreted as `redshift-averaged' quantities.

\subsection{Inferred Properties of the SMBH Binary Populations}
\label{sec:results_astro_constraints}

\begin{figure}
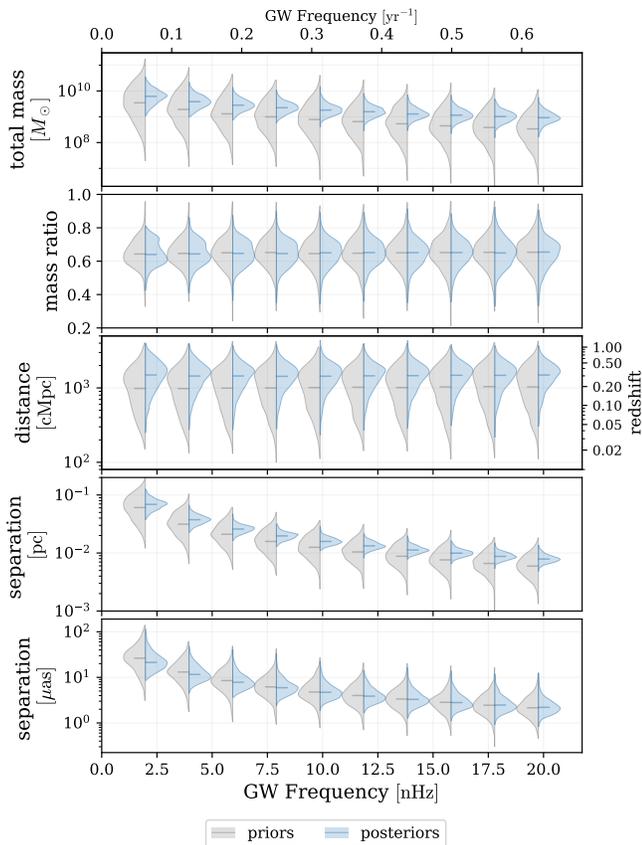

    \includegraphics[width=1\columnwidth]{{{fig-11_params-by-freq}}}
    \caption{Distributions of GWB-weighted binary parameters versus GW frequency for populations drawn from the \libphenomuniform{} library priors (left-hand violins in gray), as well as from the posteriors after fitting to the first five frequency bins of the 15 yr NANOGrav data (right-hand violins in blue).  For each population sample, the $\hc^2$-weighted averages of each parameter are calculated, and the distributions of those averages are plotted here. The GWB favors high total binary masses, especially in the lowest frequency bins, and typical binary separations range from $\sim 10^{-1}$ - $10^{-2}$ pc between 2 and 20 nHz.}
    \label{fig:binary_weighted_params}
\end{figure}

While a large amount of information is encoded in the GWB spectra, there are numerous degeneracies -- particularly in the current low signal-to-noise regime.  For example, given a particular GWB spectral shape, a certain GWB amplitude can be produced by a large number of lower-mass SMBH binaries, or a small number of higher-mass SMBH binaries.  To determine the characteristic properties of SMBH binaries contributing to the GWB, we calculate the distribution of GW-weighted average binary parameters.  We use 1000 draws from the posteriors of the \libphenomuniform{} fits.  For each draw, the $\hc^2$-weighted parameters are calculated over 100 realizations.  This gives a distribution of average parameters for each draw and each realization, which are plotted in Figure~\ref{fig:binary_weighted_params}. As in all of our analysis, we fit binary population models to only the lowest five frequency bins in the 15 yr data set. However, in order to better visualize the trends in binary population parameters with GW frequency, Figure~\ref{fig:binary_weighted_params} shows the \libphenomuniform{} library priors and posteriors for ten frequency bins.

The GWB is characterized by the most massive SMBH binaries in the Universe with $M \gtrsim 10^{8.5} \, \msol$, and extending to just above $10^{10} \, \msol$ at the lowest frequency bins.  At higher frequencies, as binaries evolve more quickly and fewer binaries occupy each frequency bin, these most massive systems become rarer and the typical masses decrease. Because of the trend in mass, the typical separations of binaries decreases more rapidly than $f^{-2/3}$ as would be expected for a fixed mass.  The binary total masses are the most strongly constrained parameters when comparing between the library priors and the posteriors.  This is unsurprising given (a) the strong dependence of GW strain of binary mass, and (b) the numerous varying model parameters that affect the masses.  The mass-ratio distributions, on the other hand, are nearly constant across the band, and narrowly localized for both the priors and posteriors.  Typical binary mass ratios are almost entirely above $q \sim 0.5$.  Note that this is determined primarily by our fiducial parameters for mass-ratio dependence in the GPF and GMT\footnote{We do allow the GPF mass-ratio dependence ($\gamma_{p,0}$) to vary in the \libgwonlyext{} and \libphenomext{} libraries (see \S~\ref{sec:app_constrained_priors}).  The parameter posteriors are virtually identical to the priors, suggesting that varying the mass-ratio dependence has little effect on the goodness of fit.}.  For the latter in particular, the GMT scales as $q^{-1}$, which strongly disfavors extreme mass-ratio pairs.

\begin{figure*}
    \includegraphics[width=\textwidth]{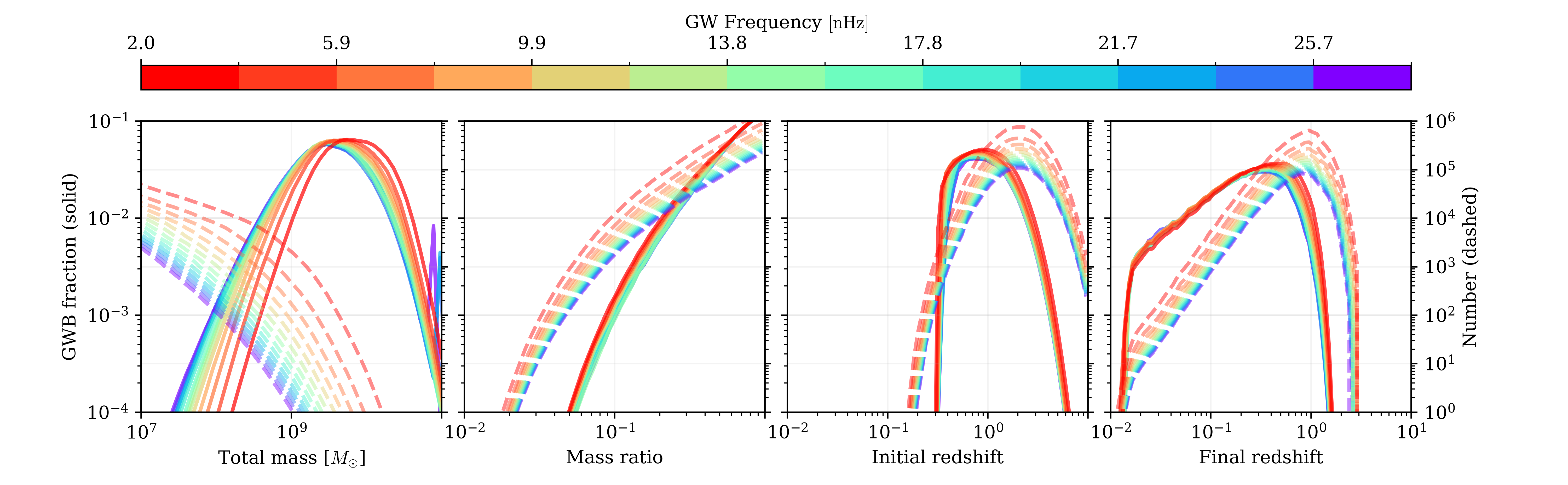}
    \caption{Comparison of binary parameters by their fractional contribution to the GWB ($\hc^2$; solid), and their total number of binaries (dashed).  Each line is colored by the GW frequency at which the binaries emit.
    The third panel ($z_\tr{init}$), corresponds to the `initial' galaxy merger redshift, while the fourth panel ($z_\tr{final}$) is the redshift at which the binary is emitting GWs.  This figure demonstrates that the GW signal is produced by a relatively small and highly biased sub-sample of a much larger population.}
    \label{fig:gwbfracvsnum}
\end{figure*}

Across the PTA band, the binaries producing the GWB are typically at many hundreds to a few thousands of comoving Mpc ($z\approx 0.15$ - $0.9$).  Average redshift posteriors are higher than the priors due to fits preferring shorter binary lifetimes.  Binary separations are tightly constrained by the strong constraints on the binary masses. Typical separations are just below $10^{-1}\, \tr{pc}$ at the lowest frequency bin ($\approx 2$ nHz), down to just below $10^{-2} \, \tr{pc}$ at the tenth frequency bin ($\approx 20$ nHz).  Projecting these separations at the cosmological distances of the binaries leads to angular separations of tens of $\mu\tr{as}$.

\begin{figure*}
    \includegraphics[width=\textwidth]{{{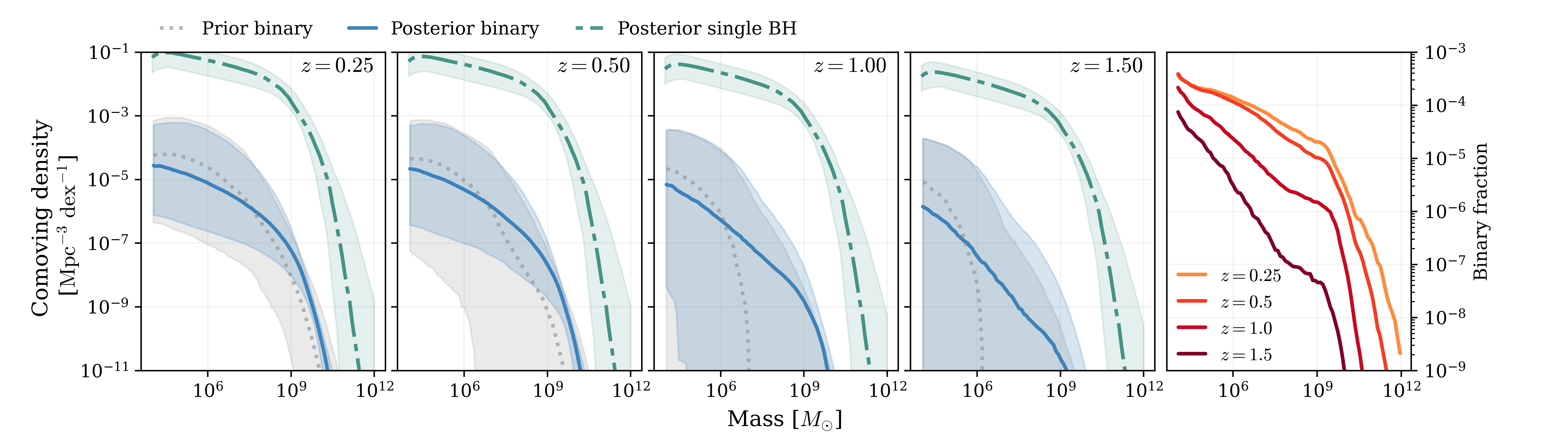}}}
    \caption{Left four panels show comoving number density per logarithmic interval of mass.  Densities are shown for binary black holes in the NANOGrav frequency band at the redshift indicated in the upper-right panel for the priors (grey dotted line) and the posteriors (blue solid line) derived from our \libphenom{} population models.  Densities are also shown for the total SMBH population (teal dash-dotted line).  Shaded regions show the 68\% distributions.  The rightmost panel shows the binary fraction as a function of mass for the four selected redshifts.
    The implied mass functions of our posteriors prefer relatively high density of black holes larger than $M=10^{9}\ \msol$, and our posteriors are overall more confined than our priors at lower redshifts.  }
    \label{fig:implied_smbhb_function}
\end{figure*}

Having explored the distributions of GWB-weighted binary properties from our fiducial model (Figure~\ref{fig:binary_weighted_params}), we now examine which binary parameter ranges contribute most to the GWB signal.  In Figure~\ref{fig:gwbfracvsnum}, we show the fraction of the GWB contributed by different portions of the binary population. Using the same posteriors as were sampled previously, we plot the fraction of $h_c^2$ that is contributed by binaries at each binary parameter value (solid lines).  The results are also separated by GW frequency (colors).  We compare these GWB fractions to the total number of binaries in our simulated populations that are emitting at a given frequency (dashed lines).  There is a stark difference between the number of binaries emitting at each frequency as a function of $M_{\mathrm{tot}}$ and their relative contribution to the GWB signal.  Lower-mass binaries are far more common, but the signal is dominated by the rare, massive black holes near $M_{\mathrm{tot}} = 10^{9.5}$--$10^{10}\ \msol$.  For example, binaries between $M_{\mathrm{tot}} = 10^{9.2}$--$10^{10.4}\ \msol$ make up $2.6 \times 10^{-4}$ of all binaries emitting in the lowest frequency band, but they make up 73\% of the signal at that frequency. Similarly, we are preferentially sensitive to the largest mass ratios.

Although the bulk of the GWB signal is made up of binaries at $z > 0.4$, it is most sensitive to the nearest binaries, relative to the underlying SMBH binary population.  Since, all else being equal, the binaries that come from later galaxy mergers (lower $z_{\rm init}$) will enter the PTA band at later times (lower $z_{\rm final}$), the GWB source population is biased towards these lower-redshift sources.  These biases affect all frequencies relatively uniformly, with the mass bias slightly more pronounced at lower frequencies.  This reflects the increased total number of binaries at low frequencies, which allows for the rarer, higher-mass SMBH binaries to dominate, as also seen in Figure~\ref{fig:binary_weighted_params}.

Note the distinction between Figure~\ref{fig:binary_weighted_params}, which shows $\hc^2-$weighted binary parameter distributions, and Figure~\ref{fig:gwbfracvsnum}, which shows the fractional contribution to the GWB of binaries with given parameters. These figures are very closely related, but Figure~\ref{fig:binary_weighted_params} shows the representative properties of GWB binaries, while Figure~\ref{fig:gwbfracvsnum} shows the fractional GWB contribution of actual binaries.

Our simulated populations also contain individually loud, high-mass binaries that can contribute substantially to the GWB.  These sources, apparent in Figure~\ref{fig:gwbfracvsnum} as spikes in the GWB fraction at high total masses, are likely the types of sources that will be detectable as continuous-wave signals by PTAs \citep{2023arXiv230103608A, nanograv_15yr_cw}.  While they can occur at a range of frequencies, they are typically expected mid-band ($f \sim 3 - 30 \, \tr{nHz}$) where the overall GWB amplitude has dropped somewhat but a sizeable population of binaries remains \citep{Kelley+2018, Becsy+2022}.

\begin{figure}
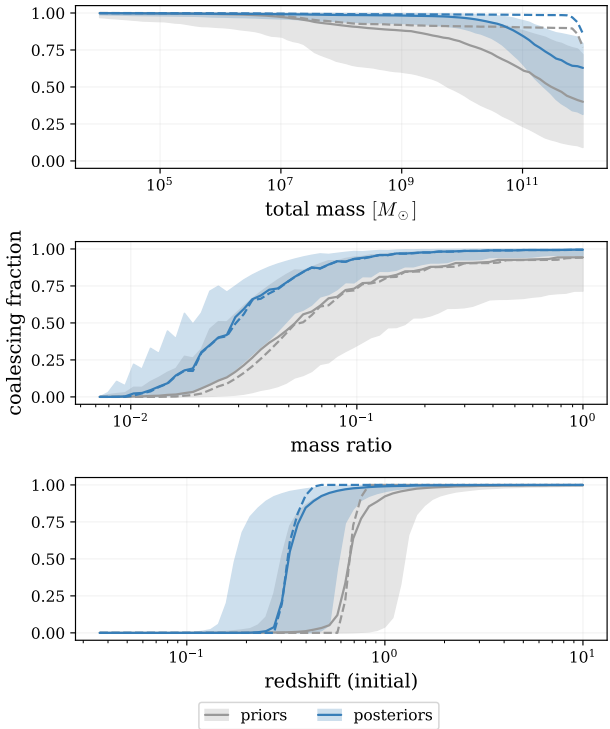

    \includegraphics[width=1\columnwidth]{{{fig-14_coal-fracs}}}
    \caption{Fraction of binary systems that reach the PTA band before redshift zero in our simulated populations.  The {\em top} panel shows the total binary mass $M$, the middle panel shows the binary mass ratio $q$, and the bottom panel shows the initial binary redshift (defined as the galaxy merger redshift) $z_{\rm init}$. Solid lines show median values over 1000 samples from the priors (grey) and posteriors (blue) after fitting to the 15 yr data.  The shaded regions correspond to the $50\%$ inter-quartile range.  The dashed lines correspond to binary subsets, with $q > 0.2$ in the \textit{top} panel, with $M > 10^8 \, \msol$ in the \textit{middle} panel, and both cuts in the \textit{bottom} panel.  Fits to the 15 yr data strongly favor shorter binary lifetimes, allowing larger fractions of binaries to coalesce.  The nearly step-function behavior in the bottom panel marks the redshift at which the look-back time matches the sample's binary-lifetime parameter.
    }
    \label{fig:coalescing_fraction}
\end{figure}

The dashed lines in the leftmost panel of Figure~\ref{fig:gwbfracvsnum} indicate the general shape of the mass function of binaries contributing to the GWB. In Figure \ref{fig:implied_smbhb_function}, we explicitly calculate the implied mass function of SMBH binaries at frequencies within the PTA band for several redshifts ($z_{\rm final}$).  Distributions are shown for populations drawn from a sample of the \libphenomuniform{} parameter space priors and from our 15 yr spectral fit posteriors.  For comparison, we also show the mass functions for non-binary SMBHs at each redshift, and in the rightmost panel, we show the fraction of SMBHs at each redshift that are in binaries. At $z \geq 1.0$, the implied mass function is consistent with a wide range of values, indicating weak PTA constraints on high-redshift SMBH binary populations. This partly reflects the steep drop in binary fraction with increasing redshift; binaries that emit in PTA bands are $\sim 10$ -- $100$ times rarer at $z=1.5$ than at $z=0.25$.

Because of this, the binary number density increases over time (note that the {\em non}-binary number density also increases with time as required for monotonic SMBH growth, but the evolution of the non-binary mass function is slight compared to the binary mass function evolution). The shape of the binary mass function also significantly evolves, such that at lower redshifts we see a much clearer turnover at $M \sim 10^9 \msol$ that more closely traces the shape of the non-binary mass function. The binary mass function is also much more tightly constrained at low redshifts than at high redshifts, especially at the high-mass end. These $\gtrsim 10^9 \msol$ binaries at $z<1$ are precisely the objects that comprise the bulk of the GWB (Figure \ref{fig:gwbfracvsnum}).

Figures~\ref{fig:binary_weighted_params} -- \ref{fig:implied_smbhb_function} examine the properties of binaries that emit in PTA bands. In Figure~\ref{fig:coalescing_fraction}, we show the fraction of {\em all} binaries that reach the lowest frequency bin of the 15 yr data before redshift zero.  Although binaries in the PTA bands will not reach coalescence on human timescales, the binary lifetime from PTA frequencies to merger is significantly shorter than a Hubble time.\footnote{For the binary masses characteristic of the GWB (e.g., Figure~\ref{fig:binary_weighted_params}), the inspiral time from the lowest frequency bin ranges from roughly $0.05 - 50$~Myr.} We therefore use the fraction of binaries reaching PTA frequencies as a proxy for the fraction of systems that coalesce entirely before redshift zero.

Whether a given binary coalesces or stalls is determined by its formation redshift combined with the binary evolution time to reach the PTA band.  Fitting to the 15 yr data strongly favors short binary lifetimes, which drives the difference between priors and posteriors.  The redshift (\textit{bottom} panel) at which the coalescing fraction reaches zero marks the redshift at which the look-back time of the Universe matches the binary lifetime of the model.  The median posterior value of this lifetime is $\approx2.8$~Gyr, corresponding to a redshift of $z\approx0.25$---where the median coalescing fraction reaches zero.  The gradual increase of coalescing fraction with redshift after this point is due to the additional delays from the GMTs.  Coalescing fractions reach unity once the combined binary lifetimes and GMTs are longer than the look-back time.

The gradual increase in coalescing fraction with mass ratio is due primarily to the GMT's strong dependence on mass ratio ($\tgal\sim q_\star^{-1}$) in the fiducial model.  The more gentle decline in coalescing fraction at the highest total masses is due to more extreme mass-ratio systems, as seen in the comparison between dashed lines ($q > 0.2$) and solid lines (all $q$).  This is caused by a combination of increased GMTs, and also the increased binary coalescing times within the PTA band---produced by more extreme mass-ratio systems that have longer GW-driven inspiral times.

We compare the \libphenom{} model to a higher-dimensional library (\libphenomext{}) which includes variations in two of the GMT parameters, as well as others, in Appendix~\ref{sec:app_constrained_priors} and Figure~\ref{fig:high-dim_astro-2}.  Generally, all of the recovered posteriors are consistent between the different libraries for parameters that they have in common, suggesting that our choices of fiducial parameters are sufficiently representative of the binary evolution parameter space.  The posteriors on the additional parameters themselves are generally broad.  The exception is the GMT parameters which, like the phenomenological evolution parameters, strongly favor shorter lifetimes as a way of producing higher GWB amplitudes.


\section{Discussion}
\label{sec:discussion}

The NANOGrav PTA has detected a common-spectrum correlated stochastic process that is consistent with an astrophysical GWB.  In our 15 yr data set \citepalias{nanograv_15yr_dataset, nanograv_15yr_detchar}, we find evidence of the Hellings-Downs correlations that would definitively mark this signal as GW in origin \citepalias{nanograv_15yr_gwb}.  In this paper, we have presented analysis of the NANOGrav 15 yr data set under the assumption that these data represent a GWB produced by SMBH binaries.  With reasonable choices of astrophysical parameters governing galaxy masses, galaxy mergers, SMBH masses, and SMBH binary inspiral timescales, we are able to reproduce the inferred GWB amplitude and spectral shape.  We find that the data are suggestive of a GWB spectral turnover at low frequencies, as expected for binary inspiral driven by astrophysical environments. However, the broad free-spectrum posteriors from the 15 yr data are still consistent with the canonical $\alpha = 2/3$ ($\gamma = 13/3$) power law expected for GW-driven inspiral.

Figure~\ref{fig:gwb_predictions} compares the posteriors for the GWB amplitude and spectral index inferred from the 15 yr data with a wide variety of GWB model predictions in the literature (see also Table \ref{Table:gwb_predictions}).  Although the inferred GWB amplitudes are within the range of some of these model predictions, they lie at the high-amplitude end of this range. The implied GWB amplitude from the NANOGrav data therefore indicates that SMBH binary model parameters differ from standard expected values, although still remaining within reasonable bounds.

In this analysis, we have generated simulated populations of SMBH binaries and GWB spectra and fit them to the observed 15 yr signal.  Our fiducial models explore a six-dimensional space of binary evolution parameters. Relative to typically assumed values for these parameters, our results indicate that the inferred GWB amplitude could be achieved with short binary hardening timescales, higher galaxy number densities (translating to higher galaxy merger rates), or higher normalization of the \mmbulge{} relation. This may be accomplished if multiple parameters differ somewhat from standard expectations, or if a small number of parameters differ more significantly.

Our models also demonstrate that the GWB signal is strongly dominated by the most massive, high-mass-ratio SMBH binaries, even among the subset of SMBH binaries emitting in PTA bands. The binaries contributing to the GWB form at typical redshifts of $z \approx 0.15$ - $0.9$. Their typical separations (which are tightly constrained via the SMBH masses) range from $\sim 0.1$ - $0.01$ parsec; this corresponds to binary angular separations of tens of $\mu$as. Owing to the short binary lifetimes preferred by the 15 yr data set, most of these binaries will merge by $z=0$---the coalescing fraction is near unity for binaries that form by $z \sim 0.25$. In addition, we note that our simulated binary populations contain loud, high-mass continuous-wave sources that could be detected above the GWB.

Because we are currently in the low signal-to-noise regime of GWB observations, we are still limited in our ability to make stringent parametric constraints.  In this analysis, our constraints on the binary population inferred from the GWB spectral shape and amplitude are dependent on both our choices of priors and on which 15 yr GWB measurements are used.  NANOGrav continues to collect data from an ever increasing number of pulsars.  43 pulsars were included in the 12.5-year\footnote{See \citet{NANOGrav12p5_data_nb} for the complete 12.5-year data set.} analysis \citep{NANOGrav12p5_background}, versus 67 pulsars in the current 15 yr analysis.  As of summer 2023, we are timing roughly 75 pulsars with a total baseline of over 17 years.  Also, NANOGrav data are currently being combined with those from other PTAs to create a new IPTA data set that will contain over 100 pulsars \citep{IPTA_2022}.  These efforts will improve our GWB measurement accuracy, along with our ability to constrain SMBH binary physics.  The theoretical forecasts from \citet{Pol+2021_astro4cast}, for example, suggest that analyses using the future NANOGrav 20-year data set would be far more constraining than the 15 yr data.  In that case, the authors found that subtle differences in the degree of environmental coupling could be distinguished based on spectra with nearly identical reference amplitudes, but differing low-frequency spectral shapes.

While we are unable to definitively attribute the inferred GWB signal to SMBH binaries at the current signal-to-noise, we show that all of the signal's features are consistent with binaries.  Nonetheless, many other possible origins of the GWB have been proposed, as detailed in \citetalias{nanograv_15yr_new_physics}.  It is worth emphasizing that SMBH binaries must necessarily form throughout the Universe as a natural product of galaxy mergers. If the inferred GWB is not dominantly produced by SMBH binaries, the lack of GW signal from inspiraling SMBH binaries must somehow be accounted for. One possibility is that SMBH binaries usually stall outside of the GW-driven regime of inspiral, which could occur if gas- and stellar-driven processes are insufficient to bring binaries to the GW-dominated regime. If indeed the so-called ``final parsec problem" \citep[e.g.,][]{BBR1980} lengthens most inspiral timescales to a Hubble time or longer, the resulting GWB from binaries could be attenuated to amplitudes well below the inferred 15 yr signal.  Even in this pessimistic case, multiple studies have suggested that triple-SMBH interactions would still produce a detectable GWB signal \citep{Volonteri+2003, Hoffman2007, Ryu_2018, Bonetti2018b}.  In either case, this would also imply the existence of a large population of stalled SMBH binaries in the local Universe.

Additional data is required to resolve the origin of the GWB.  One of the strongest distinguishing features between different source models is the significantly higher degree of anisotropy for binaries as opposed to new physical processes \citepalias{nanograv_15yr_new_physics}.  From the 15 yr data set, the first limits on anisotropy have now been placed \citep{nanograv_15yr_anisotropy}.  While the limits are still consistent with astrophysical expectations for binary populations \citep[e.g.][]{Sato-Polito+Kamionkowski-2023}, they will become significantly more constraining over time \citep{Ali-Haimoud+2021, Pol+2022_anisotropy}.

Eventually, individual `continuous-wave' GW sources will also become distinguishable above the GWB, if it is indeed produced by binaries.  Different models have produced a variety of expectations for the plausibility of continuous-wave source detection in the near future \citep{Sesana+2009, Rosado_2015, Mingarelli+2017, Kelley+2018, Becsy+2022}.  A search for continuous-wave sources has yielded improved upper limits on their occurrence rates in both the 12.5-year \citep{12p5-CW} and 15 yr data \citep{nanograv_15yr_cw}.  A continuous-wave detection would present the exciting possibility of multi-messenger detections: GWs from a single SMBH binary for which an electromagnetic counterpart could be identified.  Such a multi-messenger source would provide a wealth of information about the origin of low-frequency GWs, the astrophysical environment of SMBH binaries, and SMBH accretion processes \citep{2019BAAS...51c.490K}.

In the next decade, the Laser Interferometer Space Antenna (LISA) will begin operation in the $\sim$millihertz band, sensitive to the merger of SMBH binaries with masses in a range between $\sim 10^4 - 10^8$ $\msol$ out to $z\sim10$ \citep{LISA:2022yao}.  While LISA promises to reveal the elusive formation mechanism of massive black-hole seeds in the early Universe, this range of binary masses is more poorly constrained and far more challenging to model than the higher-mass PTA binaries.  The approaches and analyses developed for studying PTA GW sources will be crucial for paving the way for LISA science.  The identification of LISA electromagnetic counterparts will also be much more difficult, further motivating the development of techniques for PTA sources and generalizing them to signals at lower masses.

If the GWB signal is indeed produced by astrophysical binaries, it will be the first proof that SMBH binaries do indeed form, evolve to sub-parsec separations, and eventually coalesce.  These systems will join the new landscape of multimessenger GW astrophysics, offering the opportunity to study the most extreme and energetic environments in the Universe and to probe the closely coupled co-evolution of galaxies and their nuclear engines.  If instead the GWB has a different cosmological origin, it may provide answers to the most outstanding questions in fundamental physics that challenge the standard model and $\Lambda$CDM.  In either case, PTAs have cracked open the era of low-frequency GW astronomy.


\section*{Acknowledgments}


\emph{Author contributions.}
An alphabetical-order author list was used for this paper in recognition of the fact that a large, decade timescale project such as NANOGrav is necessarily the result of the work of many people. All authors contributed to the activities of the NANOGrav collaboration leading to the work presented here, and reviewed the manuscript, text, and figures prior to the paper's submission. 
Additional specific contributions to this paper are as follows.
%
L.Z.K.~proposed and coordinated the project.  Development of the \holodeck{} population modeling framework was led by L.Z.K., with contributions from A.C-C., D.W., E.C.G., J.M.W., K.G., M.S.S., \& S.C.  Libraries were generated by J.S., K.G., L.B., L.Z.K., \& N.G-D.  Development of the Gaussian process modules was led by J.S., with contributions from D.W. \& J.M.W.  Development of the fitting infrastructure was led by W.G.L., with contributions from J.S.  Gaussian process training was performed by D.W. \& J.S., with diagnostics performed by J.M.W.  The 15 yr GWB data products were provided by the NANOGrav Detection Working Group.  The analysis was performed by J.S., K.G., \& L.Z.K., who produced most of the figures, and by L.B.  Additional analysis code and interpretation were provided by D.J.D. and E.C.G.  The collection of literature predictions, and production of the literature review figure, was performed by M.C.  The paper was written by J.C.R., J.S., K.G., L.B., L.Z.K., \& M.C.  A.Mi, B.B., D.J.D., J.S.H., M.T.L., M.V., S.J.V., S.R.T., T.J.W.L., \& T.D. provided feedback on the analysis and manuscript.

\emph{Acknowledgments.}
The NANOGrav collaboration receives support from National Science Foundation (NSF) Physics Frontiers Center award numbers 1430284 and 2020265, the Gordon and Betty Moore Foundation, NSF AccelNet award number 2114721, an NSERC Discovery Grant, and CIFAR. NANOGrav is part of the International Pulsar Timing Array (IPTA).

The Arecibo Observatory is a facility of the NSF operated under cooperative agreement (AST-1744119) by the University of Central Florida (UCF) in alliance with Universidad Ana G. M{\'e}ndez (UAGM) and Yang Enterprises (YEI), Inc. The Green Bank Observatory is a facility of the NSF operated under cooperative agreement by Associated Universities, Inc. The National Radio Astronomy Observatory is a facility of the NSF operated under cooperative agreement by Associated Universities, Inc.

This work was conducted in part using the resources of the Advanced Computing Center for Research and Education (ACCRE) at Vanderbilt University, Nashville, TN.  This work received computational support from UCF’s Advanced Research Computing cluster, operated by the University of Central Florida.  This research used the Savio computational cluster resource provided by the Berkeley Research Computing program at the University of California, Berkeley (supported by the UC Berkeley Chancellor, Vice Chancellor for Research, and Chief Information Officer).
This research was supported in part through computational resources and services provided by Advanced Research Computing at the University of Michigan, Ann Arbor.
This work was also conducted in part using the Thorny Flat HPC Cluster at West Virginia University (WVU), which is funded in part by National Science Foundation (NSF) Major Research Instrumentation Program (MRI) Award number 1726534, and West Virginia University.
This work utilized the Alpine high performance computing resource at the University of Colorado Boulder. Alpine is jointly funded by the University of Colorado Boulder, the University of Colorado Anschutz, Colorado State University, and the National Science Foundation (award 2201538). This work utilized the Blanca condo computing resource at the University of Colorado Boulder. Blanca is jointly funded by computing users and the University of Colorado Boulder.

L.B. acknowledges support from the National Science Foundation under award AST-1909933 and from the Research Corporation for Science Advancement under Cottrell Scholar Award No. 27553.
P.R.B. is supported by the Science and Technology Facilities Council, grant number ST/W000946/1.
S.B. gratefully acknowledges the support of a Sloan Fellowship, and the support of NSF under award \#1815664.
The work of R.B., R.C., D.D., N.La., X.S., J.P.S., and J.T. is partly supported by the George and Hannah Bolinger Memorial Fund in the College of Science at Oregon State University.
M.C., P.P., J.C.R. and S.R.T. acknowledge support from NSF AST-2007993.
M.C. and N.S.P. were supported by the Vanderbilt Initiative in Data Intensive Astrophysics (VIDA) Fellowship.
K.Ch., A.D.J., and M.V. acknowledge support from the Caltech and Jet Propulsion Laboratory President's and Director's Research and Development Fund.
K.Ch. and A.D.J. acknowledge support from the Sloan Foundation.
Support for this work was provided by the NSF through the Grote Reber Fellowship Program administered by Associated Universities, Inc./National Radio Astronomy Observatory.
Support for H.T.C. is provided by NASA through the NASA Hubble Fellowship Program grant \#HST-HF2-51453.001 awarded by the Space Telescope Science Institute, which is operated by the Association of Universities for Research in Astronomy, Inc., for NASA, under contract NAS5-26555.
K.Cr. is supported by a UBC Four Year Fellowship (6456).
M.E.D. acknowledges support from the Naval Research Laboratory by NASA under contract S-15633Y.
T.D. and M.T.L. are supported by an NSF Astronomy and Astrophysics Grant (AAG) award number 2009468.
E.C.F. is supported by NASA under award number 80GSFC21M0002.
G.E.F., S.C.S., and S.J.V. are supported by NSF award PHY-2011772.
K.A.G. and S.R.T. acknowledge support from an NSF CAREER award \#2146016.
The Flatiron Institute is supported by the Simons Foundation.
S.H. is supported by the National Science Foundation Graduate Research Fellowship under Grant No. DGE-1745301.
N.La. acknowledges the support from Larry W. Martin and Joyce B. O'Neill Endowed Fellowship in the College of Science at Oregon State University.
Part of this research was carried out at the Jet Propulsion Laboratory, California Institute of Technology, under a contract with the National Aeronautics and Space Administration (80NM0018D0004).
D.R.L. and M.A.Mc. are supported by NSF \#1458952.
M.A.Mc. is supported by NSF \#2009425.
C.M.F.M. was supported in part by the National Science Foundation under Grants No. NSF PHY-1748958 and AST-2106552.
A.Mi. is supported by the Deutsche Forschungsgemeinschaft under Germany's Excellence Strategy - EXC 2121 Quantum Universe - 390833306.
P.N. acknowledges support from the BHI, funded by grants from the John Templeton Foundation and the Gordon and Betty Moore Foundation.
The Dunlap Institute is funded by an endowment established by the David Dunlap family and the University of Toronto.
K.D.O. was supported in part by NSF Grant No. 2207267.
T.T.P. acknowledges support from the Extragalactic Astrophysics Research Group at E\"{o}tv\"{o}s Lor\'{a}nd University, funded by the E\"{o}tv\"{o}s Lor\'{a}nd Research Network (ELKH), which was used during the development of this research.
S.M.R. and I.H.S. are CIFAR Fellows.
Portions of this work performed at NRL were supported by ONR 6.1 basic research funding.
J.D.R. also acknowledges support from start-up funds from Texas Tech University.
J.S. is supported by an NSF Astronomy and Astrophysics Postdoctoral Fellowship under award AST-2202388, and acknowledges previous support by the NSF under award 1847938.
C.U. acknowledges support from BGU (Kreitman fellowship), and the Council for Higher Education and Israel Academy of Sciences and Humanities (Excellence fellowship).
C.A.W. acknowledges support from CIERA, the Adler Planetarium, and the Brinson Foundation through a CIERA-Adler postdoctoral fellowship.
O.Y. is supported by the National Science Foundation Graduate Research Fellowship under Grant No. DGE-2139292.
D.J.D. received funding from the European Union's Horizon 2020 research and innovation programme under Marie Sklodowska-Curie grant agreement No. 101029157, and from the Danish Independent Research Fund through Sapere Aude Starting Grant No. 121587.

\facilities{Arecibo, GBT, VLA}

\software{
    acor,
    astropy \citep{2018AJ....156..123A},
    ceffyl \citep{Lamb+2023},
    chainconsumer \citep{Hinton2016},
    cython \citep{cython2011},
    emcee \citep{emcee2013},
    enterprise \citep{enterprise},
    enterprise\_extensions \citep{enterpriseext},
    George \citep{george2015},
    holodeck \citep[][in prep]{holodeck},
    jupyter \citep{jupyter2016},
    kalepy \citep{kalepy},
    matplotlib \citep{matplotlib2007},
    numba \citep{numba2015},
    numpy \citep{numpy2011},
    ptmcmc \citep{PTMCMCSampler},
    scipy \citep{2020SciPy-NMeth},
}

\newpage
\appendix

\renewcommand{\thetable}{\Alph{section}\arabic{table}}
\renewcommand{\thefigure}{\Alph{section}\arabic{figure}}
\setcounter{table}{0}
\setcounter{figure}{0}


\section{GWB Predictions in the Literature}
\label{sec:app_lit_models}

\begin{figure*}
    \includegraphics[width=\textwidth]{{{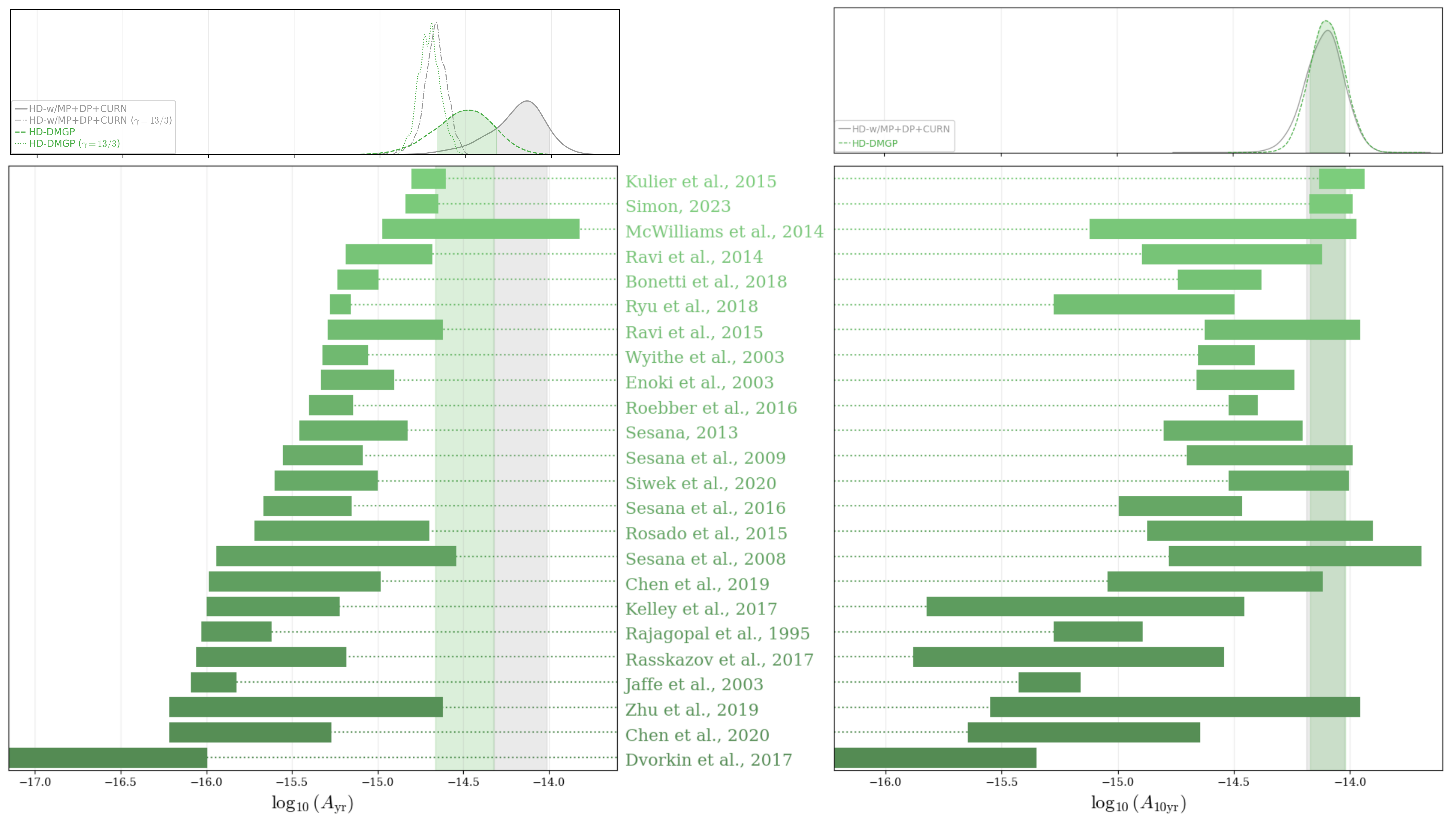}}}
    \caption{Literature predictions for the amplitudes $A_{\rm yr}$ and $A_{\rm 10yr}$ of the GWB at frequencies of $\lr{1 \, \yr}^{-1}$ and $\lr{10 \, \yr}^{-1}$, respectively, compared to the NANOGrav 15 yr results.  While it is most common to reference GWB amplitudes at $f=1\,\pyr$, NANOGrav constraints are primarily derived at lower frequencies and thus the $f=\lr{10 \, \yr}^{-1}$ values are much more representative of current PTA constraints.  The green horizontal bars indicate the $16^{\rm th}$ - $84^{\rm th}$ percentile uncertainty regions for each prediction; these model predictions are also listed in Table~\ref{Table:gwb_predictions}. The amplitude distributions in the upper panels correspond to the posterior probability distributions of GWB amplitude for power-law models fit to the GWB free-spectrum posteriors (\hdall---gray solid curves, and \hdgp---green dashed curves; see Figure 5 in \citetalias{nanograv_15yr_gwb}). The $\gamma=13/3$ slice is also shown for the $A_{\rm yr}$ values in dashed-dotted and dotted lines for the \hdall and \hdgp values respectively, since those are the most directly comparable to many of the models included here. The green and gray shaded regions are the corresponding 68\% credible intervals. The 15 yr NANOGrav results are within the bounds of some model predictions but require GWB amplitudes at the higher end of the predicted ranges.
    }
    \label{fig:gwb_predictions}
\end{figure*}

In Figure~\ref{fig:gwb_predictions} and Table~\ref{Table:gwb_predictions}, we summarize model predictions of GWB amplitudes from the literature. For each reference, amplitude, predictions are cited at frequencies of $\lr{1 \, \yr}^{-1}$ and $\lr{10 \, \yr}^{-1}$.  While it is most common to reference GWB amplitudes at $f=1\,\pyr$, NANOGrav constraints are derived primarily at much lower frequencies.  Small deviations in power-law indices at low frequencies can lead to large changes in the corresponding $A_\tr{yr}$ amplitudes.  The amplitudes at $f=\lr{10 \, \yr}^{-1}$ are much more representative of current PTA constraints.  Figure~\ref{fig:gwb_predictions} also shows the posterior distributions for the inferred GWB amplitudes obtained from the 15 yr data NANOGrav when assuming a power-law model; results are shown for both the \hdall\ and the \hdgp\ models. Numerous literature model predictions overlap with the inferred GWB amplitudes from the 15 yr data, especially when comparing the amplitude at $f=\lr{10 \, \yr}^{-1}$. Even so, the 15 yr results lie at the higher-amplitude end of the predicted ranges.

\begin{table}
\centering
\begin{tabular}{r|c|c|c|c}
Model & $A_{\rm yr} \; 16^{\rm th}$ &  $A_{\rm yr} \; 84^{\rm th}$ &  $A_{\rm 10yr} \; 16^{\rm th}$ &  $A_{\rm10yr} \; 84^{\rm th}$ \\
\hline
     \citet{Rajagopal_1995}& 9.32E$-$17 & 2.41E$-$16 & 5.31E$-$16 & 1.28E$-$15 \\
     \citet{Jaffe_2003} & 8.10E$-$17 & 1.50E$-$16 & 3.75E$-$16 & 6.93E$-$16\\
     \citet{Wyithe_2003} & 4.77E$-$16 & 8.84E$-$16 & 2.22E$-$15 & 3.89E$-$15 \\
     \citet{Enoki_2004}& 4.70E$-$16 & 1.25E$-$15 & 2.18E$-$15 & 5.77E$-$15 \\
     \citet{Sesana+2008} & 1.15E$-$16 & 2.88E$-$15 & 1.66E$-$15 & 2.04E$-$14 \\
     \citet{Sesana_2009} & 2.79E$-$16 & 8.21E$-$16 & 1.98E$-$15 & 1.03E$-$14 \\
     \citet{Sesana_2013a} & 3.50E$-$16 & 1.50E$-$15 & 1.58E$-$15 & 6.26E$-$15 \\
     \citet{McWilliams_2014}& 1.07E$-$15 & 1.51E$-$14 & 7.58E$-$16 & 1.07E$-$14\\
     \citet{Ravi_2014} &  6.51E$-$16 & 2.10E$-$15 & 1.27E$-$15 & 7.62E$-$15 \\
     \citet{Kulier_2015} & 1.58E$-$15 & 2.51E$-$15 & 7.36E$-$15 & 1.16E$-$14 \\
     \citet{Ravi_2015} & 5.10E$-$16 & 2.40E$-$15 & 2.37E$-$15 & 1.11E$-$14 \\
     \citet{Rosado_2015} & 1.91E$-$16 & 2.01E$-$15 & 1.34E$-$15 & 1.26E$-$14 \\
     \citet{Roebber_2016} & 4.00E$-$16 & 7.23E$-$16 & 3.00E$-$15 & 4.00E$-$15 \\
     \citet{Sesana_2016} & 2.15E$-$16 & 7.08E$-$16 & 1.01E$-$15 & 3.43E$-$15 \\
     \citet{Rasskazov_2017} & 8.74E$-$17 & 6.57E$-$16 & 1.32E$-$16 & 2.87E$-$15\\
     \citet{Dvorkin_2017} &8.74E$-$17 & 6.57E$-$16 & 1.32E$-$16 & 2.87E$-$15 \\
     \citet{Kelley+2017b} &  1.00E$-$16 & 6.00E$-$16 & 1.50E$-$16 & 3.50E$-$15 \\
     \citet{Ryu_2018} &  5.30E$-$16 & 7.00E$-$16 & 5.30E$-$16 & 3.20E$-$15 \\
     \citet{Bonetti2018b} & 5.83E$-$16 & 1.01E$-$15 & 1.81E$-$15 & 4.18E$-$15 \\
     \citet{Zhu_2019} & 6.10E$-$17 & 2.40E$-$15 & 2.83E$-$16 & 1.11E$-$14\\
     \citet{Chen+2019} & 1.04E$-$16 & 1.05E$-$15 & 9.02E$-$16 & 7.63E$-$15\\
     \citet{Chen_2020} &  6.10E$-$17 & 5.40E$-$16 & 2.26E$-$16 & 2.27E$-$15 \\
     \citet{Siwek+2020} & 2.50E$-$16 & 1.00E$-$15 & 3.00E$-$15 & 9.94E$-$15\\
     \citet{Simon-2023} & 1.46E$-$15 & 2.26E$-$15 & 6.67E$-$15 &  1.03E$-$14

\end{tabular}
\caption{Literature predictions for the amplitudes $A_{\rm yr}$ and $A_{\rm 10yr}$ of the GWB at frequencies of $\lr{1 \, \yr}^{-1}$ and $\lr{10 \, \yr}^{-1}$, respectively.  Column (1) gives the literature reference, columns (2) \& (3) give the $16^{\rm th}$ \& $84^{\rm th}$ percentiles for the uncertainty region of the corresponding prediction of $A_{\rm yr}$, and columns (4) \& (5) give the $16^{\rm th}$ \& $84^{\rm th}$ percentiles for  $A_{\rm 10yr}$ predictions. Figure \ref{fig:gwb_predictions} provides a visual comparison of these predictions to the NANOGrav 15 yr results.}
\label{Table:gwb_predictions}
\end{table}

\setcounter{table}{0}
\setcounter{figure}{0}
\section{Semi-Analytic Model Parameterizations}
\label{sec:app_sam_params}
In \S~\ref{sec:meth_binary_pops}, we presented the semi-analytic models and the underlying equations used in this paper. In Table~\ref{Table:sam_params}, we detail all of the model parameters, showing the fiducial values for fixed parameters as well as the prior distributions for parameters that are varied when fitting the models to the 15 yr data. These models are summarized Table~\ref{Table:models}.

\begin{table}
\centering
\begin{tabular}{c|c|c|c|c}
model component & symbol & fiducial value   & uniform priors    & astrophysical priors \\
\hline
\multirow{6}{*}{GSMF$^{a}$ $\lr{\gsmffunc}$}
    & $\psi_0$          & -                   & $\mathcal{U}(-3.5, \, -1.5)$& $\mathcal{N}(-2.56, \, 0.4)$ \\
    & $\psi_z$          & $-0.60$             & -                           & -                            \\
    & $m_{\psi,0}$      & -                   & $\mathcal{U}(10.5, \, 12.5)$& $\mathcal{N}(10.9, \, 0.4)$  \\
    & $m_{\psi,z}$      & $+0.11$             & -                           & -                            \\
    & $\alpha_{\psi,0}$ & $-1.21$             & -                           & $\mathcal{N}(-1.2, \, 0.2)$  \\
    & $\alpha_{\psi,z}$ & $-0.03$             & -                           & -                            \\
\hline
\multirow{7}{*}{GPF$^{b}$ $\lr{P}$}
    & $P_0$             & $+0.033$            & -                           & -                        \\
    & $\alpha_{p,0}$    & $0.0$               & -                           & -                        \\
    & $\alpha_{p,z}$    & $0.0$               & -                           & -                        \\
    & $\beta_{p,0}$     & $+1.0$              & -                           & $\mathcal{N}(0.8, 0.4)$  \\
    & $\beta_{p,z}$     & $0.0$               & -                           & -                        \\
    & $\gamma_{p,0}$    & $0.0$               & -                           & $\mathcal{N}(0.5, 0.3)$  \\
    & $\gamma_{p,z}$    & $0.0$               & -                           & -                        \\
\hline
\multirow{7}{*}{GMT$^{c}$ $\lr{\tgal}$}
    & $T_0$             & $+0.5 \, \tr{Gyr}$  & -                           & $\mathcal{U}(0.2, 5.0)$ Gyr     \\
    & $\alpha_{t,0}$    & $0.0$               & -                           & -                               \\
    & $\alpha_{t,z}$    & $0.0$               & -                           & -                               \\
    & $\beta_{t,0}$     & $-0.5$              & -                           & $\mathcal{U}(-2.0, 0.0)$        \\
    & $\beta_{t,z}$     & $0.0$               & -                           & -                               \\
    & $\gamma_{t,0}$    & $-1.0$              & -                           & -                               \\
    & $\gamma_{t,z}$    & $0.0$               & -                           & -                               \\
\hline
\multirow{3}{*}{\mmbulge{}$^{d}$ $\lr{\mmbulgefunc}$}
    & $\mmbamp$         & -                   & $\mathcal{U}(7.6, \, 9.0)$     & $\mathcal{N}(8.6, \, 0.2)$          \\
    & $\mmbplaw$        & $+1.10$             & -                              & $\mathcal{N}(1.2, \, 0.2)$          \\
    & $\mmbscatter$     & -                   & $\mathcal{U}(0.0, \, 0.9)$ dex & $\mathcal{N}(0.32, \, 0.15)$ dex    \\
    & $\mmbulgefbulge$  & $+0.615$            & -                              & -                                   \\
\hline
\multirow{5}{*}{phenom $\lr{\frac{da}{dt}}$}
    & $\tlifetime$          & -             & $\mathcal{U}(0.1, \, 11.0)$ Gyr  & $\mathcal{U}(0.1, \, 11.0)$ Gyr  \\
    & $\hardrchar$          & $+10^2$ pc    & -                                & -                                \\
    & $\hardainit$          & $+10^3$ pc    & -                                & -                                \\
    & $\hardnuinner$        & -             & $\mathcal{U}(-1.5, \, 0.0)$      & $\mathcal{U}(-1.5, \, +0.5)$     \\
    & $\hardnuouter$        & $+2.5$        & -                                & -

\end{tabular}
\\\footnotesize{$^a$Fiducial GSMF values are based on \citet{Chen+2019}, while the `astrophysical library' parameters are based on fits to the data from \citet{Tomczak+2014}. $^b$GPF parameters are based on a comparison of \citet{Conselice+2003}, \citet{Bluck+2012}, \citet{Mundy+2017}~\&~\citet{Duncan+2019}.  $^c$GMT parameters are based on a comparison of \citet{Conselice+2008}, \citet{Boylan-Kolchin+2008}, \citet{Conselice-2009}, and \citet{Snyder+2017}.  $^{d}$\mmbulge{} parameters are based on \citet{Gultekin+2009}, \citet{KH13}, and \citet{MM13}, with bulge fractions based on \citet{Lang+2014} and \citet{Bluck+2014}.}
\caption{Astrophysical parameters of our semi-analytic population models. Units are denoted where relevant; all other parameters are defined to be dimensionless. For libraries, we denote uniform distributions with $\mathcal{U}(\tr{min}, \tr{max})$ and normal distributions with $\mathcal{N}(\tr{mean}, \tr{std.~dev.})$.}
\label{Table:sam_params}
\end{table}

\begin{table}
\centering
\begin{tabular}{c|c|c|c|c|c|c}
\hline
\multirow{2}{*}{model name} & \multicolumn{5}{c|}{parameters varied (by model component)} & \multirow{2}{*}{priors}  \\
 & GSMF & GPF & GMT & \mmbulge{} & phenom & \\ \hline
{\bf \libphenomuniform} & $\gsmfnorm$, $\gsmfmass$ & - & - & $\mmbamp$, $\mmbscatter$ & $\tlifetime$, $\hardnuinner$ & uniform \\ \hline
\libphenomastro & $\gsmfnorm$, $\gsmfmass$ & - & - & $\mmbamp$, $\mmbscatter$ & $\tlifetime$, $\hardnuinner$ & astrophysical \\
\libgwonlyuniform & $\gsmfnorm$, $\gsmfmass$ & - & - & $\mmbamp$, $\mmbscatter$ & - & uniform \\
\libphenomextastro & $\gsmfnorm$, $\gsmfmass$, $\alpha_{\psi,0}$ & $\beta_{p,0}$, $\gamma_{p,0}$ & $T_0$, $\beta_{t,0}$ & $\mmbamp$, $\alpha_\mu$, $\mmbscatter$ & $\tlifetime$, $\hardnuinner$ & astrophysical \\
\libgwonlyextastro & $\gsmfnorm$, $\gsmfmass$, $\alpha_{\psi,0}$ & $\beta_{p,0}$, $\gamma_{p,0}$ & $T_0$, $\beta_{t,0}$ & $\mmbamp$, $\alpha_\mu$, $\mmbscatter$ & - & astrophysical \\
\hline
\end{tabular}
\caption{Summary of semi-analytic SMBH binary population models used in this work. Model parameters are defined in \S~\ref{sec:meth_binary_pops}, and their fiducial values and assumed prior distributions are given in Table~\ref{Table:sam_params}. The first model (\libphenomuniform), indicated in boldface, is what we refer to as our fiducial model. Throughout the text, when the assumed priors can be omitted from the model name without loss of clarity, we simply refer to the models as \libphenom{}, \libgwonly{}, \libphenomext{}, and \libgwonlyext{}. The latter two (also referred to as the ``extended models") are discussed in Appendix~\ref{sec:app_constrained_priors}.}
\label{Table:models}
\end{table}

\setcounter{table}{0}
\setcounter{figure}{0}
\section{Higher-dimensional parameter spaces}
\label{sec:app_constrained_priors}

We have shown that in our fiducial, six-dimensional \libphenom{} parameter space, a wide range of semi-analytic model parameters are consistent with current measurements of the GWB.  In this library, a large number of additional parameters are held fixed to astrophysically motivated values.  When fitting to the 15 yr GWB data, uniform priors on the included parameters are typically used.  We have compared these results to fits of the same parameter space, but adopting more informed priors based on the astronomical literature.

Currently it is not feasible to run MCMC fits using Gaussian processes that have been trained to significantly larger parameter spaces.  However, we have generated higher dimensional libraries and directly evaluated them against the 15 yr GWB spectra at the library grid points themselves.  In this way we can weight the input parameters by the resulting likelihoods to obtain posteriors without using MCMC to dynamically explore the domain.  This approach allows us to examine the effects of freeing additional parameters.

\begin{figure*}
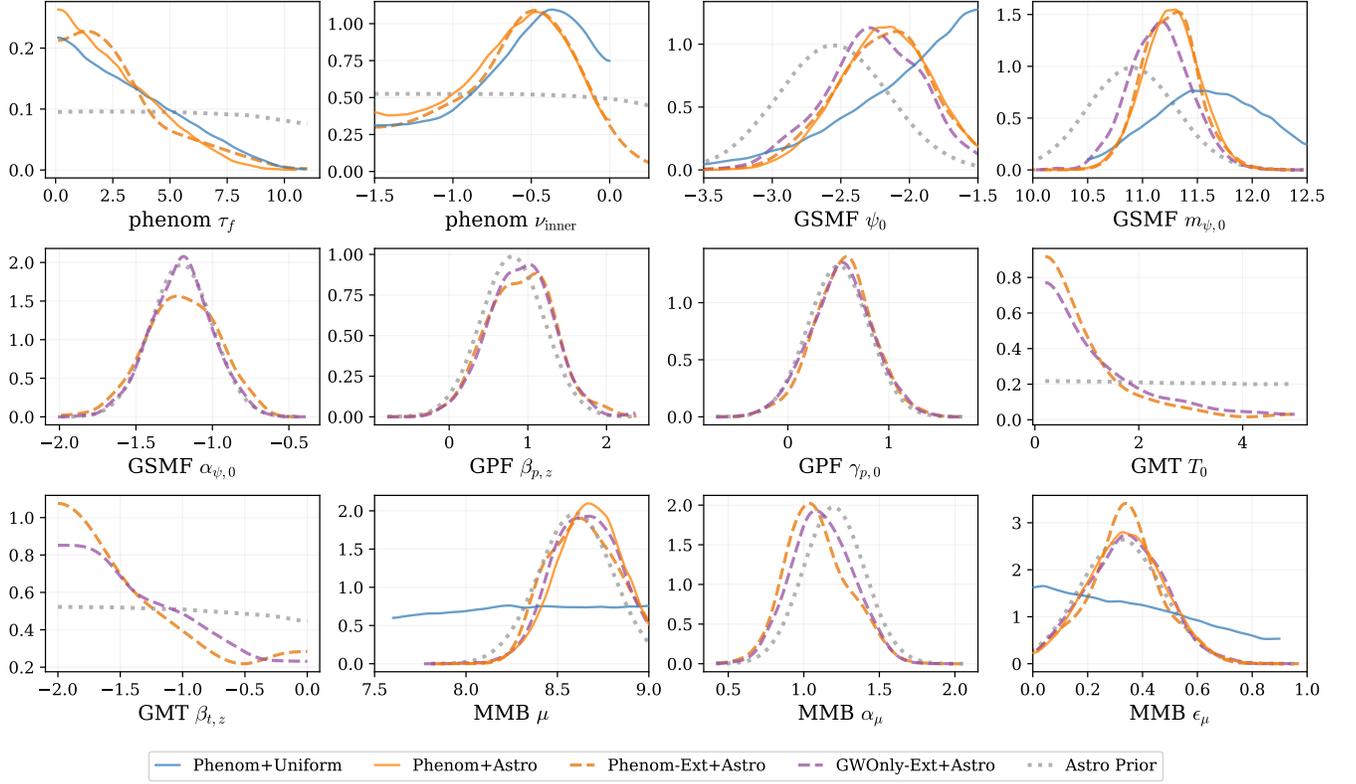

    \includegraphics[width=1\columnwidth]{{{fig-c1_ext-libs}}}
    \caption{Semi-Analytic Model posteriors comparing our fiducial six-dimensional \libphenom{} model (solid) to our much larger \libphenomext{} parameter space (dashed).  The standard version of the \libphenom{} library with uniform priors (blue) is also compared to a version which is fit against the 15 yr GWB spectra using the same astrophysically motivated priors from the \libphenomastro{} distributions (orange).  The priors are also shown for the astrophysically motivated case (grey).
    }
    \label{fig:high-dim_astro-2}
\end{figure*}

Figure~\ref{fig:high-dim_astro-2} compares the parameter posteriors for our fiducial library against the larger parameter spaces, including a twelve-dimensional phenomenological version (\libphenomext{}, orange dashed), and ten-dimensional GW-only version (\libgwonlyext{}, purple dashed).  Blue solid lines show the standard \libphenom{} fits with uniform priors, while orange solid lines show the same six-dimensional library, but fitting with priors taken from the \libphenomext{} distributions.

Posteriors from the six-dimensional phenomenological models are entirely consistent with the ten- \& twelve-dimensional \libphenomext{} libraries when using the same priors.  Setting the additional parameters to fixed values does not bias the resulting measurements, nor does it lead to under-estimating the width of posterior distributions.  This is likely the case because the current 15 yr NANOGrav spectral measurements include large uncertainties.  As the data improves, it will become more important to fully explore the parameter space.

As previously discussed, for many parameters the shape of the priors does significantly impact the recovered posteriors.  This is particularly noticeable in the GSMF parameters ($\gsmfnorm$, and $\gsmfmass$), and the \mmbulge{} parameters ($\mmbamp$, and $\mmbscatter$) where there is significant degeneracy between parameters that broadly change the amplitude of the GWB spectrum.

\bibliography{refs}{}
\bibliographystyle{aasjournal}

\end{document}